\documentstyle[aps,prl,multicol,epsf,epsfig]{revtex}
\def\CC{{\rm\kern.24em \vrule width.04em height1.46ex depth-.07ex
\kern-.30em C}}
\def\P{{\rm I\kern-.25em P}}
\def\RR{{\rm
         \vrule width.04em height1.58ex depth-.0ex
         \kern-.04em R}}

\def\RR{{\rm\kern.24em \vrule width.04em height1.46ex depth-.07ex
\kern-.30em R}}
\def\P{{\rm I\kern-.25em P}}
\def\RR{{\rm
         \vrule width.04em height1.58ex depth-.0ex
         \kern-.04em R}}

\newcommand{\PHI}{\mbox{\boldmath${\phi}$}}
\newcommand{\U}{{\cal U}}
\newcommand{\cp}{{\bf CP}}
\newcommand{\be}{\begin{equation}}
\newcommand{\ee}{\end{equation}}
\newcommand{\bq}{\begin{eqnarray}}
\newcommand{\eq}{\end{eqnarray}}

\newcommand{\Sp}{\,\,\,\,\,\,\,\,\,\,\,\,\,}
\newcommand{\no}{\nonumber\\}

\newcommand{\p}{\partial}
\newcommand{\la}{\lambda}

\newcommand{\D}{{\cal D}}

\newcommand{\al}{\alpha}
\newcommand{\bi}{\beta}
\newcommand{\ga}{\gamma}

\newcommand{\si}{\sigma}

\newcommand{\th}{\theta}

\newcommand{\TH}{\mbox{\boldmath${\theta}$}}

\newcommand{\lan}{\langle}
\newcommand{\ran}{\rangle}
\newcommand{\A}{{\cal A}}

\begin{document}
\draft
\title{Quantum Holonomies for Quantum Computing }
\author{Jiannis Pachos$^{1,2}$ and Paolo Zanardi $^{2,3}$ }
\address{ 
$^1$ Max Planck Institut f\"ur Quantenoptik, D-85748 Garching, Germany
\footnote{Present address.}\\
$^2$ Institute for Scientific Interchange (ISI) Foundation,
 Viale Settimio Severo 65, I-10133 Torino, Italy\\
$^3$ Istituto Nazionale Fisica della Materia (INFM)
}
\maketitle
\begin{abstract}
{ Holonomic Quantum Computation (HQC) is an all-geometrical approach
to quantum information processing. In the HQC strategy information 
is encoded in degenerate eigen-spaces of a parametric family of 
Hamiltonians. The computational network of unitary quantum gates is 
realized by driving adiabatically the Hamiltonian parameters along 
loops in a control manifold. By properly designing such loops
the non-trivial curvature of the underlying bundle geometry gives 
rise to unitary transformations i.e., holonomies that implement
the desired unitary transformations. Conditions necessary for universal 
QC are stated in terms of the curvature associated to the non-abelian 
gauge potential (connection) over the control manifold. In view of 
their geometrical nature the holonomic gates are robust against several 
kind of perturbations and imperfections. This fact along with the 
adiabatic fashion in which gates are performed makes in principle HQC 
an appealing way towards universal fault-tolerant QC.
}
\end{abstract}
\date{today}
\pacs{PACS numbers: 03.67.Lx, 03.65.Bz}
\maketitle
\begin{multicols}{2}
\narrowtext
\section{Introduction}

It is by-now a generally accepted fact that the laws of quantum theory 
provide in principle a radically novel, and more powerful 
way to process information with respect to 
any conceivable classically operating device \cite{QC}. 
In the last few years a big deal of activity has been devoted to devise and to implement schemes
for taking actual advantage from this extra quantum power.
In particular in Quantum Computation 
the states of a quantum system $S$ are used for encoding information in such a way 
that the final state, obtained by the unitary time-evolution of $S,$ encodes
the solution of a given computational problem. 
A system $S$ with state-space $\cal H$ (the Quantum Computer) supports universal QC if any
unitary transformation $U\in{\cal U}({\cal H})$ can be approximated with arbitrarily high accuracy
by a sequence (the {\em network}) of simple unitaries (the {\em gates}) that the experimenter is supposed 
to be able to implement. 
The case in which $S$ is a multi-partite system is the most important 
one as it allows for {\em entanglement}, a unique quantum feature that is 
generally believed to be one of the crucial elements from which polynomial or 
exponential speed-up occurs \cite{jozsa}. 

In the above picture of QC the realization of the quantum network is achieved at the physical level by turning on and off 
external fields coupled to $S$ as well as {\em local} interactions among the subsystems of $S.$ 
In other words the experimenter ``owns'' a basic set of time-dependent Hamiltonians that she/he activates at will
to perform the suitable sequences of quantum logic gates.
At variance with such a standard {\em dynamical} view of QC
more recently several authors considered {\em geometrical} and 
{\em topological} approaches \cite{KIT,FREE,AK,LLO}.
The peculiarity of these proposals is somehow striking: 
over the manifold $\cal C$ of quantum codewords one can have
a trivial Hamiltonian e.g., $H|_{\cal C}=0,$ nevertheless,
obtaining a non-trivial quantum evolution
due to the existence of an underlying geometrical/topological {\em global} structure.
The quantum gates - or a part of them - 
are then realized in terms of operations having a purely geometrical/topological nature.
Besides being conceptually intriguing on their own,
these schemes have some built-in fault-tolerant features. This latter attractive characteristic stems 
from the fact that some topological as well as geometrical quantities are inherently stable against local perturbations.
This in turn allows for Quantum Information processing inherently stable against special classes
of computational errors. 

In this paper we shall give a detailed account of {\em Holonomic Quantum Computation} (HQC) 
introduced in Ref. \cite{hol1} and further developed in Refs. \cite{hol2}, \cite{hol3}, \cite{JAMS} and 
\cite{Fujii}.
In this novel gauge-theoretic framework one is supposed to be able to 
control a set of parameters $\lambda \in {\cal M}$,
on which depends an iso-degenerate family $\cal F$ of quantum Hamiltonians $\{ H(\lambda)\}.$ 
Information is encoded in an $n$-dimensional eigen-space $\cal C$ of a distinguished $H(\lambda_0)\in{\cal F}$.
Universal QC \cite{UG} over $\cal C$ can be then obtained by adiabatically driving the control parameters along
 suitable loops $\gamma$ rooted at $\lambda_0.$
The key physical ingredient is provided by the appearance in such quantum evolutions of non-Abelian geometric contributions \cite{WIZE}
$U_\gamma\in U(n)\,(n>1)$ given by {\em holonomies} associated with a $u(n)$-valued gauge potential $A$ \cite{WY,NAK}.
In other words quantum computation in the HQC approach is nothing but 
the parallel transport of states in $\cal C$ realized by the {\em connection} $A$. 
Therefore the computational power in the HQC approach relies on the non-triviality 
of the geometry of the bundle of eigen-spaces of $\cal F$ over the manifold of control parameters, $\cal M$.
It must be noticed that also in Refs. \cite{AK,LLO}, 
even though the evolutions are not adiabatic, holonomies play an important role.
Nevertheless the connections involved in those approaches are {\em Abelian} giving rise
to $U(1)$-valued holonomies e.g., Berry phases \cite{SHWI}.
It is then clear that
in order to achieve universality these geometric gates must be supplemented 
by standard dynamical operations. On the other hand in the HQC approach {\em the whole quantum network is built
by means of holonomies}. In this sense HQC is {\em fully} geometrical.
It is worth observing that the computational subspace $\cal C$ 
can be thought of as the lowest-energy manifold of a highly symmetric quantum system;
from this point of view HQC is a kind of {\em ground-state} computation. 
This last remark points out the potential existence of a fault-tolerant \cite{Preskill} 
feature of HQC due to energy gaps and even spontaneous relaxation mechanisms.
Further fault-tolerant characteristics, of HQC models we considered, are related to the fact 
that the holonomies $U_\gamma$ realizing quantum computations turn out to depend just on the 
{\em areas} of the surfaces that the generating loops $\gamma$ span on certain two-dimensional sub-manifolds.
When this area is given one can consider even very large i.e., ``far'' from the identity
deformations of $\gamma$, but as long as they are area-preserving no errors are induced.
Moreover as far as the adiabaticity condition holds $U_\gamma$ does not depend 
on the rate at which the control loops are driven. Hence, even with respect the issue of timing, HQC
is robust.

The paper is organized as follows. In Section II the general theory of holonomic
quantum evolution is presented as well as its application to quantum information processing.
A new pedagogical proof of the adiabatic theorem is given and the properties
of the holonomies are analyzed, facilitating their application to quantum computing.
In Section III the $\cp^n$ model is presented as a theoretical holonomic arena allowing
for the analytic evaluation of the connection $A$, its field strength $F$
and a complete set of calculated holonomies. The mathematical steps which enable 
such a calculation are given in detail providing an analytical method for
calculating the holonomies (Wilson loops) much used in many areas of theoretical physics.
In Section IV a physical model is presented based on quantum optics.
Known optical devices as displacers, squeezers and interferometers are employed
as control devices performing coherent evolution of the laser photon states.
The laser beams propagate in a Kerr medium which provides the desired
degeneracy. Further similar applications to harmonic oscillator setups are
discussed, which are experimentally more viable. 
In the Appendix a more mathematical approach to the Holonomic evolutions 
is presented.

\section{General Theory}

In this section 
the general theoretical framework of HQC is reviewed.
While the exposition relies partly on Refs. \cite{hol1,hol2} some proofs have
been added which clarify the physical concept of holonomic evolutions making
this subject more approachable to the quantum information community.
 
\subsection{Quantum Evolutions}

Let us suppose that we have at disposal a family $\cal F$
of Hamiltonians that we can turn on and off in order 
to let an $N$-dimensional quantum system to evolve in a controllable way.
Formally, we assume ${\cal F}:=\{ H(\lambda)\}_{\lambda\in{\cal M}}$ 
to be a continuous family of Hermitian operators over the state-space ${\cal H}\cong \CC^N.$
The parameters $\lambda$ on which the elements of $\cal F$ depend 
will be referred to as {\em control} parameters and their manifold $\cal M$
as the {\em control manifold}, thought to be embedded in $\RR^{N^2}.$
Indeed, one has
\bq
H(\lambda)=i\,\sum_{a=1}^{N^2} \Phi_a(\lambda) \,T_a\in u(N)
\nonumber
\eq
where the $T_a$'s constitute a basis of the
$N^2$-dimensional Lie-algebra $u(N)$ of anti-hermitian matrices,
and
$\Phi\colon {\cal M}\mapsto u(N)$ is a smooth mapping that
associates to any $ \lambda$ in the control manifold
a vector in $u(N)$ with $T$-components $(\Phi_1(\lambda),\ldots,\Phi_{N^2}(\lambda)).$

The evolution of the quantum system is thought of as actively driven
 by the parameters $\lambda,$ over which the experimenter is assumed to have direct 
access and controllability.
Suppose we are able 
to drive by a {\em dynamical control process } the parameter configuration $\lambda\in{\cal M}$ through a
{ path} $\gamma\colon [0,\,T]\rightarrow {\cal M}.$
Hence, a one-parameter i.e., time-dependent family 
\be
{\cal F}_\gamma:=\{H(t):=H[\Phi\circ\gamma (t)]\colon t\in [0,\,T]\}\subset {\cal F}
\label{1-family}
\ee
is defined.
Notice that even the converse is true: any smooth family $\{H(t)\}_{t\in[0,\,T]}$
defines a path in ${\cal M}={\RR}^{N^2}.$
The quantum evolution associated to the time-dependent family (\ref{1-family})
is described by the time-dependent Schr\"odinger equation
$i\,\partial_t|\psi(t)\rangle= H(t)\,|\psi(t)\rangle$
and hence it has the operator form
\begin{equation}
U_\gamma:={\bf T}\,\exp\{ -i\,\int_0^T\! dt\, H(t)\}\in U(N)
\label{evolution}
\end{equation}
where $\bf T$ denotes chronological ordering. 
The time-dependent quantum evolution (\ref{evolution}), for a given map $\Phi,$ 
depends in general on the path $\gamma$ and {\em not} just
on the curve $\gamma([0,\,T])$ i.e., the image of $\gamma$ in the control manifold.
In other words the unitary transformation (\ref{evolution})
contains a {\em dynamical} as well as a {\em geometrical} contribution,
the former depends even on the {\em rate} at which $\gamma([0,\,T])$
is traveled along whereas the latter depends
merely on the geometrical characteristics of the curve.

From the physical point of view the parameters $\lambda$
represent in general external fields and, for multi-partite systems, couplings among the various subsystems.
To illustrate this point let us consider ${\cal H}:=(\CC^2)^{\otimes\,N}\cong \CC^{2^N}$
i.e., a $N$-qubit system.
Then a basis for $u(2^N)$ is provided by
the tensor products $T_\alpha:=\otimes_{i=1}^N \hat \sigma_{\alpha_i}$ where
$\alpha\colon\{1,\ldots,N\}\mapsto \{0,1,2,3\}$ and
$\hat \sigma_0:=\openone,\, \hat \sigma_1:=\sigma_x,\, \hat \sigma_2:=\sigma_y,\, \hat \sigma_3:=\sigma_z$
are the Pauli matrices.
It is then clear that any $\alpha$ which takes a non-zero value more than once e.g., $\alpha_i\,\alpha_j\neq 0$
describes a non-trivial interaction which generates entanglement between the qubits $i$ and $j$.
Therefore the ability to manipulate the weight of the contribution of $T_\alpha$'s
in the decomposition of $H(\lambda),$ amounts to the capacity of dynamically
controlling many-body couplings.
This goal is, of course, even conceptually more difficult to achieve
than the control of the real external fields, namely the interaction 
associated to single subsystem generators $T_\alpha.$
Finally, we stress that there is still another possibility;
the control parameters $\lambda$ could represent on {\em their own}
quantum-degrees of freedom e.g., nuclear coordinates 
in the adiabatic approximation for molecular systems, 
treated in some quasi-classical fashion.
This situation arises when one performs an adiabatic decoupling
between ``fast'' and ``slow'' degrees of freedom,
getting for the former a Hamiltonian that depends parametrically on the latter
\cite{SHWI}.
In this case the control manifold $\cal M$ is nothing but the classical configuration
manifold associated with a quantum system. 

Within this framework
the requirements for implementing universal QC \cite{UG}
can be expressed in terms of the availability of paths.
Universality is the experimental capability of driving the control parameters along
a minimal set $\{\gamma_i\}_{i=1}^g$ of paths which generate 
the basic unitary transformations $U_{\gamma_i}$'s, i.e. the gates. By sufficiency of this set we mean the
ability to approximate any $U\in U(N)$ with arbitrarily high accuracy by means of 
path sequences.

\subsection{Holonomies \label{Holo}}%%%%%%%%%%%%%%%%%%%%%%%%%%%%%%%%%%%%%%%%%%%%%%%%%%%%%%%%%%

Now we recall some basic facts about quantum holonomies. A more mathematical approach 
can be found in Appendix A, where some by-now standard material has been collected
aiming to make the paper as much as possible self-contained.

The non-Abelian holonomies are a natural generalization of the Abelian
Berry phases. 
We first assume that $\cal F$ 
is an {\em iso-degenerate} Hamiltonian family i.e., 
all the elements of $\cal F$ have 
the same degeneracy structure.
This means that a generic Hamiltonian of $\cal F$ can be written as 
$ H(\lambda)=\sum_{l=1}^R \varepsilon_l(\lambda) \, \Pi_l(\lambda)$
where $\Pi_l(\lambda)$ denotes the projector over the eigen-space
${\cal H}_l(\lambda) :=\mbox{span}\{|\psi^\alpha_{l}(\lambda)\rangle\}_{\alpha=1}^{n_l},$
with eigenvalues $\varepsilon_l(\lambda)$,
whose dimension $n_l$ is independent on the control parameter $\lambda.$
In order to preserve the $R$ degeneracies $n_l$
we also assume that over $\cal M$ there is no level-crossing i.e.,
$l\neq l^\prime \Rightarrow\varepsilon_l(\lambda)\neq \varepsilon_{l^\prime} (\lambda),
\forall \lambda\in{\cal M}.$
In addition, we shall restrict to {\em loops} $\gamma$ in the control manifold i.e.,
maps $\gamma\colon [0,\,T]\mapsto {\cal M}$ such that $\gamma(0)=\gamma(T).$ 
These conditions in the dynamics of the system and in the control
manipulations will facilitate the generation of holonomic unitaries.

Let us state the main result \cite{WIZE} on which the HQC relies.
Consider a system with the above characteristics.
When its control parameters are driven adiabatically i.e., slow with respect 
to any time-scale associated to the system dynamics, along a 
loop $\gamma$ in $\cal M$ 
any initially prepared state $|\psi_{in}\rangle\in{\cal H}$
will be mapped after the period $T$ onto the state
\begin{equation}
|\psi_{out}\rangle=
U_\gamma\,|\psi_{in}\rangle,\, U_\gamma=\oplus_{l=1}^R e^{i\,\phi_l}\,\Gamma_{A_l}(\gamma),
\label{out}
\end{equation}
where, $\phi_l:=\int_0^T d\tau\, \varepsilon_l(\lambda_\tau),$ is the dynamical phase 
whereas the matrices $\Gamma_{A_l}(\gamma)$'s represent the geometrical contributions.
They are unitary mappings of ${\cal H}_l$ onto itself
and they can be expressed by the following path ordered integrals 
\begin{equation}
\Gamma_{A_l}(\gamma) :={\bf{P}}\exp \oint_\gamma A_{l} \in U(n_l) \,\, ,\,\,\,
l=1,\ldots,R \,\, .
\label{Hol}
\end{equation}
These are the {\em holonomies} associated with the loop $\gamma,$
and the {\em adiabatic connection forms } $A_l.$ 
The latter have an explicit matrix form given by 
$A_{l} = \Pi_l(\lambda)\,d\,\Pi_l(\lambda)=\sum_\mu
A_{l,\mu}\,d\lambda_\mu,$ where \cite{SHWI} analytically
\begin{equation}
(A_{l,\mu})^{\alpha\beta}:= \langle\psi_{l}^\alpha(\lambda)|
\,{\partial}/{\partial\lambda^\mu}\,
|\psi_{l}^\beta(\lambda)\rangle
\label{conn}
\end{equation}
with $(\lambda_\mu)_{\mu=1}^d$ the local coordinates on ${\cal M}.$
The connection forms $A_l$'s are nothing but the non-Abelian gauge potentials enabling the parallel transport 
\cite{NAK} over $\cal M$ of vectors of the fiber ${\cal H}_l$.
Result (\ref{conn}) is the non-Abelian generalization
of the Berry phase connection presented first by Wilczek and Zee (1984) (see Appendix).
Due to the decomposition of the evolution operator in (\ref{out})
into distinct evolutions for each eigen-space ${\cal H}_l,$ we are able to
restrict our study to a given degenerate eigen-space with fixed $l.$

We shall present first an intuitive proof for deriving (\ref{Hol}) and (\ref{conn}),
aiming in clarifying the gauge structure interpretation of this
adiabatic evolution and in providing a more physical insight.
Without loss of generality we shall assume the family $\cal F$
to be {\em iso-spectral}. This implies that for any $\lambda\in{\cal M}$
it exists a unitary transformation $\U(\lambda)$ such that 
$H(\lambda)= \U(\lambda)\, H_0\,\U(\lambda)^\dagger,$
where $H_0:=H(\lambda_0).$ 
Upon dividing the time interval $[0,\,T]$ into $N$ equal
segments $\Delta t,$ for $\U_i=\U(\gamma(\lambda(t_i)))$ one obtains 
the evolution operator in the form
\begin{eqnarray} 
U_\gamma &= &{\bf T} e^ {-i \int_0^T \U(\lambda)\,H_0 \, \U^\dagger(\lambda)
  dt}= {\bf T} \!\!
\lim_{N\rightarrow \infty} e^ {-i \sum_{i=1}^N \U_i\,H_0 \,\U^\dagger_i \Delta
t}\nonumber \\
&= & {\bf T} \lim_{N\rightarrow \infty} \prod_{i=1}^N \U_i e^{-iH_0 \Delta t} \U_i^\dagger 
\label{evol}
\end{eqnarray} 
The third equality holds due to the smallness 
of the interval $\Delta t$ in the limit of large $N.$ The product $\U_i^\dagger \U_{i+1}$
of two successive unitaries, gives rise 
to an infinitesimal rotation of the form 
$\U_i^\dagger \U_{i+1}\approx {\openone} +\vec{A}_i \cdot \Delta \vec{\lambda}_i$, where
$\Sp ({A}_i)_\mu \equiv \U_i^\dagger {\Delta \U_i \over \Delta
(\lambda_i)_\mu}$. The connection $A$ has at 
time $t_i$ the components $(A_i)_\mu$ with $\mu=1,\ldots,d$. Hence the
evolution operator (\ref{evol}) becomes 
\be 
{\bf T} \lim_{N \rightarrow
\infty} \U_N \left( \openone -i H_0 \, N\,\cdot \Delta t + \sum_{i=1}^{N-1}
\vec{A}_i \cdot \Delta \vec{\lambda}_i \right) \U_1^\dagger. 
\label{midd}
\ee 
For the case of a closed path the initial and the final transformations $\U_1$ and $\U_N$ 
are identical as they correspond to the same point of the control parameter manifold. 
With a reparametrization they may be taken to be equal to the identity transformation. 
Now we consider an initial state $|\psi_{in}\rangle$ belonging to
an eigen-space ${\cal H}_0$ with associated eigenvalue e.g., $\varepsilon_0=0$.
Due to the time ordering symbol the actions on the state $|\psi(t)\ran$ of the 
Hamiltonian and of the connection $A$ are alternated, hence in general we cannot 
separate them into two exponentials. On the other hand, if we demand adiabaticity, 
namely very slow exchange of energy during the process, this will keep 
the state within ${\cal H}_0$, then at each time $t_i$ the state 
$|\psi(t_i)\ran$ will remain in the $\varepsilon_0=0$ energy level.
This allows to factor out in (\ref{midd}) the action of 
$H,$ thus obtaining 
\bq 
U_{\gamma}={\bf T}
\lim_{N \rightarrow \infty} \left( {\bf 1} + \sum_{i=1}^{N-1}
\vec{A}_i \cdot \Delta \vec{\lambda}_i \right) = {\bf P} \exp \oint_\gamma A \,\, ,
\nonumber
\eq 
where $A$ is projected into the subspace ${\cal H}_0.$
Notice that we replaced the time ordering with the path
ordering ${\bf P}$ as the parameter of the integration at the
last expression is the position on the loop $\gamma$.
In this proof of the non-Abelian geometrical evolution it is clear 
how the holonomy appears and which physical conditions enable its formation. 
In the same way we could have considered in addition to the equivalent 
transformations of the Hamiltonian a multiplicative function 
$\varepsilon_0(t)$ varying the energy eigenvalue. The results would be unaltered
apart from the insertion of a dynamical phase.

Let us now view some of the properties the holonomies have
in terms of gauge reparametrization
of the connection and loop composition rules.
In our context a local gauge transformation is the unitary transformation 
${\cal U}(\la)\mapsto {\cal U}(\la) g(\la)$, which
does not change the Hamiltonian operator $H_0$. Its action merely reparametrizes
the variables of the control manifold. Taking into account the properties
$gH_0=H_0g$ and $g \Pi=\Pi g$ we are able to obtain the transformation of the
connection as
$A\mapsto g^{\dagger}\,A \,g +g^{\dagger}\,dg,\,(g\in U (n))$.
It immediately follows that the holonomy transforms as 
$\Gamma_A\mapsto g^\dagger\, \Gamma_A\, g$.
Notice that in the new coordinates the state vectors $|\psi\rangle$ i.e., the sections, 
become $g^{\dagger}\,|\psi\rangle$.
This property makes it clear that the holonomy transformation has an intrinsic
i.e., coordinate-free, meaning.
Furthermore, the holonomy has the following property in terms of the loops.
We define (setting $T=1$) the loop space at a given point $\lambda_0\in{\cal M}$
as 
\bq
L_{\lambda_0}:=\{\gamma\colon [0,\,1]\mapsto {\cal M}\,/\,
\gamma(0)=\gamma(1)=\lambda_0\}
\nonumber
\eq
over a point $\lambda_0\in{\cal M}.$
Let us stress that, as far as the manifold $\cal M$ is connected, the distinguished 
point $\lambda_0$ does not play any role. 
In this space we introduce a composition law for loops 
\begin{equation}
(\gamma_2\cdot \gamma_1)(t)=\theta(
\frac{1}{2} -t)\,\gamma_1(2\,t)+
\theta(t-\frac{1}{2})\,\gamma_2(2t-1)
\label{compo}
\end{equation}
and a unity element $\gamma_0(t) \equiv \lambda_0,\,t\in[0,\,1]$
moreover with $\gamma ^{-1}$ we shall denote the loop 
$t\mapsto \gamma(1-t).$

The holonomy can be considered as a map $\Gamma_{A}\colon L_{\lambda_0}\mapsto U(n_l)$,
whose basic properties can be easily derived from eq. (\ref{Hol}):
\begin{itemize}
\item[i)] $\Gamma_A(\gamma_2\cdot\gamma_1)=\Gamma_A(\gamma_2)\,\Gamma_A(\gamma_1)$;
by composing loops in $\cal M$ one obtains
a unitary evolution that is the product of the evolutions associated with the individual loops,
\item[ii)] $\Gamma_A(\gamma_0)=\openone$;
staying at rest in the parameter space corresponds to no evolution at all,
\item[iii)] $\Gamma_A(\gamma^{-1})=\Gamma_A^{-1}(\gamma)$;
in order to get the inverse holonomy one has to traverse the path $\gamma$ with reversed orientation,
\item[iv)] $\Gamma_A(\gamma\circ \varphi)=\Gamma_A(\gamma),$ where $ \varphi$ is any
diffeomorphism of $[0,\,1]$; as long as adiabaticity holds 
the holonomy does not depend on the speed at which the path is traveled 
but just on the path geometry.
\end{itemize}

From the properties listed above it is easy to show that 
the set $\mbox{Hol}(A):=\Gamma_A(L_{\lambda_0})$ is a {\em subgroup} of $U(n).$
Such a subgroup is known as the {\em holonomy group} of the connection $A$. 
When the holonomy group coincides with the whole $U(n)$ then the connection $A$ 
is called {\em irreducible}. The notion of irreducibility plays a crucial 
role in HQC in that it corresponds to the computational notion of {\em universality} \cite{UG}.
In order to evaluate if this condition is fulfilled by a given connection it is
useful to consider the {\em curvature} $2$-form $F=\sum_{\mu\nu}F_{\mu\nu}\,dx^\mu\wedge dx^\nu$ 
associated with the $1$-form connection $A$
whose components
\begin{equation}
F_{\mu\nu} =\partial_\mu A_\nu-\partial_\nu A_\mu + [A_\mu,\,A_\nu].
\label{curvature}
\end{equation}
The relation of the curvature with irreducibility is given by the following statement \cite{NAK}:
{\em the linear span of the $F_{\mu\nu}$'s is the Lie algebra of the holonomy group}.
It follows in particular that when the $F_{\mu\nu}$'s span the whole $u(n)$
the connection is irreducible.

\subsection{Holonomic Quantum Computation}

The unitary holonomies (\ref{Hol}) are the main ingredient of our approach to QC.
From now on we shall consider a given subspace ${\cal H}_l$
(accordingly the label $l$ will be dropped).
Such a subspace, denoted by $\cal C$, will represent
our quantum {\em code}, whose elements will be the quantum information encoding
codewords.
The crucial remark \cite{hol1} is that {\em when the connection is irreducible,
for any chosen unitary transformation $U$ over the code
there exists a path $\gamma$ in $\cal M$ such that $\|\Gamma_A(\gamma)-U\|
\le \epsilon ,$ with $\epsilon$ arbitrarily small.}
This means that {\em any computation on the code $\cal C$
can be realized by adiabatically driving the control parameter configuration $\lambda$ along
a suitable closed path $\gamma.$}

In particular we aim to constructing specific logical gates by moving along their
corresponding loops. Initially, the degenerate states are prepared in to a ``ground'' state,
interpreting the $|0...0\ran$ state of $m$ qubits. 
The statement of irreducibility of the connection $A$ relates 
a particular unitary $U$ with the loop $\gamma_U$ over which the connection is
integrated to give $\Gamma_A(\gamma_U)=U$. 
Hence, there are loops in the control space such that the associated holonomies
give, for example, a one qubit Hadamard gate or a two qubit ``controlled-not'' gate. 

Let us emphasize 
the fact that one can perform {\em universal} QC by only using quantum holonomies is remarkable.
Indeed this kind of quantum evolutions is quite special,
yet it contains in a sense the full computational power.
On the other hand
one has to pay the price given by the restriction of the computational space
from $\cal H$ to its subspace $\cal C.$
Notice that,
for the irreducibility property to hold, a necessary condition is clearly given by
$d\,(d-1)/2\ge n^2$ where $d:=\mbox{dim}\,{\cal M}.$
In particular this implies that for an exponentially large code $\cal C$
one has to be able to manipulate an exponentially large number of control parameters.

Moreover like in any other scheme for QC, once the computation is completed a final state measurement is performed.
To this aim 
it could be useful to lift the energy degeneracy 
in order to be able address energetically 
the different codewords \cite{hol2}.
This can be done 
by switching on an external perturbation in a coherent fashion.

We conclude this section by discussing the {\em computational complexity} issue.
The computational subspace $\cal C$ does not have in general a tensor product structure.
This means that it cannot be viewed in a natural way as the state-space of a multi-partite
system for which the notion of quantum entanglement makes sense.
The latter, on the other hand, is known to be one of the crucial ingredients that provides to QC 
its additional power with respect to
classical computation. It follows that,
from this point of view, the scheme for HQC described so far is potentially incomplete. 
Indeed -- as it will be illustrated later by explicit examples -- if $N=$ dim ${\cal C} =2^k$
i.e., we encode in $\cal C$ $k$ qubits, then for obtaining with a multi-partite structure
a universal set of gates one needs $O(N)$ elementary holonomic loops.
Thus in general one has an {\em exponential} slow-down in computational complexity.

In Ref. \cite{hol2} we argued how one can in principle overcome such a drawback
by focusing on a class of HQC models with a multi-partite structure given from the very beginning.
The basic idea is simple: one considers an holonomic family $\cal F$ 
associated to a genuine multi-partite quantum system such that local 
(one- and two-qubit) gates can be performed by holonomies.
Then from standard universality results of QC \cite{UG} stems that 
efficient quantum computations can be performed.
An explicit example of the above strategy is formalized as follows \cite{hol2}.

Let us consider $N$ {\em qu-trits}. The state space is then given by
${\cal H}_j\cong {\CC}^3=\mbox{span}\{ |\alpha\rangle_j \, /\,\alpha=0,1,2\}$.
The holonomic (iso-spectral) family has the built-in local structure 
${\cal F} =\{ H_{ij}(\lambda_{ij})\} $ where the local Hamiltonians $H_{ij}$ have a non trivial actions 
only on the $i$-th and $j$-th factors of $\cal H.$
Moreover, $H_{ij}$ admits a four-dimensional degenerate eigen-space 
${\cal C}_{ij}:=
\mbox{span}\{ |\alpha\rangle_i \otimes |\beta\rangle_j\,/\, \alpha,\beta =0,1\}
\subset {\cal H}_i\otimes{\cal H}_j\cong {\bf{C}}^9.
$
If the $H_{ij}$'s allow for universal HQC over ${\cal C}_{ij}$ then universal QC 
can be {\em efficiently} implemented over 
\bq
{\cal C}:= \mbox{span}\{\otimes_{i=1}^N |\alpha_i\rangle_i\,/\, \alpha_i=0,1\}\cong ({\CC}^2)^{\otimes\,N}.
\nonumber
\eq

\section{The $\cp^n$ Holonomic Construction}

In this section we shall consider a theoretical model
where the holonomic ideas can be materialized. The $\cp^n$ model will be 
considered for which the Hamiltonian due to its degeneracy has such a symmetric structure
as to allow the control manifold ${\cal M}$ to be the $n$-dimensional complex 
projective space $\cp^n$. For quantum computation
we are interested in finding the particular loops which generate 
various holonomic gates and eventually constructing a complete
set of universal gates.

The path ordering prescription given in (\ref{order}) makes hard the explicit analytical
evaluation of the holonomies. In order to tackle this problem we
employ two procedures.
On the one hand we study loops restricted onto particular two dimensional 
sub-manifolds having easily calculated holonomies. 
This geometric restriction overcomes 
the difficulties connected with path ordering, allowing the
evaluation of a complete set of basic holonomies.
On the other hand it is possible to compose 
a generic unitary operator with combinations of elements within this set. 
Eventually, by the loop composition properties of the holonomies it is 
possible to find its corresponding composed loop.
Even if the techniques presently known do not
give the possibility to calculate the holonomy of the most general loop, we 
shall obtain families of loops and their corresponding holonomies from which 
any desirable group element may be constructed.

\subsection{The $\cp^n$ Model}

Consider the degenerate Hamiltonian $H_0=\varepsilon | n+1\rangle \langle n+1 |$
acting on the state-space ${\cal H}\cong \CC^{n+1}=\mbox{span} \{|\al \ran \}_{\alpha=1}^{n+1}$. 
We shall take as the family ${\cal F}$ the whole orbit 
${\cal O}(H_0):=\{ {\cal U}\,H_0\,{\cal U}^\dagger\,/\,{\cal U}\in U(n+1)\}$ of $H_0$
under the adjoint action of the unitary group $U(n+1)$. 
This orbit is isomorphic to the $n$-dimensional complex projective space
\begin{eqnarray}
{\cal O}(H_0)
\cong \frac{ U(n+1)}{U(n)\times U(1)} \cong\frac{SU(n+1)}{U(n)}
\cong
{\bf{CP}}^n.
\nonumber
\end{eqnarray}
Each point, ${\bf z}$, of the ${\bf{CP}}^n$ manifold 
corresponds to a unitary matrix
${\cal U}({\bf z}) =U_1(z_1)U_2(z_2)...$ $U_n(z_n)$, where
$U_\al(z_\al)=\exp [ G_\al(z_\al)]$ with 
$G_\al(z_\al)=z_\al |\al\ran\lan n + 1| -\bar z _\al |n ~+~ 1 \ran \lan \al |$ 
and $z_\al=\th_\al e^{i \phi_\al}$, for $\al=1,...,n$. 
We shall assume in the following the set $(\TH, \,\PHI)$ as real coordinates for $\cp^n.$
The eigen-states of the rotated Hamiltonians are 
\bq
&&
|\al(\TH,\PHI)\ran:={\cal U}(\TH, \PHI) |\al \ran =\cos \th_\al |\al \ran -
\no \no
\exp ({-i \phi_\al})&&
\sin \th_\al \sum_{j>\al} ^{n+1} \exp({i \phi_j}) \sin \th_j \prod_{j>\ga>\al} \cos
\th_\ga |j \ran
\label{state1}
\eq
and 
\bq
|n+1(\TH, \PHI) \ran := &&{\cal U}(\TH, \PHI) |n+1 \ran =
\no \no
&&
\sum^{n+1}_{j=1} \exp ({i \phi_j}) \sin
\th_j \prod_{\ga<j} \cos \th_\ga |j \ran 
\label{state2}
\eq
where we have defined $\th_{n+1}:=\pi/2$ and $\phi_{n+1}:=0$. The first $n$ ones (\ref{state1}) 
have zero eigenvalue while the last one (\ref{state2}) has eigenvalue $\varepsilon$.
Notice that for $n=1$ the standard $2$-level model with the Abelian Berry phase is recovered.

\subsection{The Connection $A$ and the Field Strength $F$}

By using definition (\ref{conn}) the components of the connection $A$ can be 
explicitly computed. Their particular form depends on the bundle of the degenerate spaces
described in (\ref{state1}) and (\ref{state2}). For $\cp^n$ $A$ has
$2n$ component as many as the dimensions of the manifold. These
$u(n)$-valued connection components over ${\bf{CP}}^n$ are anti-hermitian matrices as 
dictated by (\ref{conn}).
In detail the only non-zero elements of the matrix $A^{\th_\bi}$ ($\beta=1,\ldots,n$)
are
$A^{\th_\bi}_{\bar \al \bi}$ for $\bar \al=1,\ldots, \bi-1$, given by
\begin{eqnarray}
A^{\th_\bi}_{\bar \al \bi}&&= \lan \bar \al| {\cal U}^\dagger { \p \over \p \th_\bi} {\cal U} |\bi\ran
\no \no
&&
=e^{i(\phi_{\bar \al}- \phi_\bi)}\, \sin \th_{\bar \al} 
\prod_{\bi>\ga>\bar \al} \cos \th_\ga \,\,,
\label{niao}
\end{eqnarray}
as well as $A^{\th_\bi}_{\bar \al \bi}=-A^{\th_\bi}_{\beta \bar\alpha}$ 
which guarantees the anti-hermiticity. These are $n$ components corresponding 
to the $n$ $\theta$-coordinates of $\cp^n$. 
The anti-hermitian matrix $A^{\phi_\bi}$ corresponding to the $n$ $\phi$-coordinates 
has non-zero elements for $\al=\bi$ and $\al \geq \bar \al$ given by
\begin{eqnarray}
A^{\phi_\bi}_{\bar \al \bi}&&= \lan \bar \al| {\cal U}^\dagger { \p \over \p \phi_\bi} {\cal U} |\bi\ran
\no \no 
&&
=-i e^{i(\phi_{\bar \al}-\phi _\bi)} \sin
\th_\bi \sin \th_{\bar \al} \prod _{\bi \geq\ga > \bar \al } \cos \th_\ga,
\nonumber
\end{eqnarray}
where we assumed $\prod _{\bi \geq\ga > \bi } \cos \th_\ga=1$, 
and for $\bi>\al$ and $\al\geq \bar \al$ by
\begin{eqnarray}
A^{\phi_\bi}_{\bar \al \al} && = \lan \bar \al| {\cal U}^\dagger { \p \over \p \phi_\bi} {\cal U} |\al \ran
\no \no
&&
=i e^{i(\phi_{\bar \al}-\phi _\al)} \sin \th_{\al} \sin
\th_{\bar \al} \sin^2 \th_\bi
\!\!\! \prod _{\bi >\ga > \al }\!\!\!\! \cos \th_\ga \! \! \!
\!\prod _{\bi > \bar \ga >\bar \al } \!\! \! \! \cos \th_{\bar \ga}.
\nonumber
\end{eqnarray}

Having the transformations ${\cal U},$ which allow us to fix $A,$ we are able to 
determine also the curvature and check the irreducibility properties of the 
connection for the $\cp^n$ model.
By using the definition (\ref{curvature}) and setting $z_\alpha=z^0_\alpha+i\,z^1_\alpha,$
one finds that at ${\bf{z}}=0$ the components of the curvature
are given by 
\bq
F_{z_\alpha^i z_\beta^j}(0) =\left. \Pi_{\bf z}\,[\frac{\partial U_\alpha}
{\partial z^i_{\alpha}},\,\frac{\partial U_\beta}
{\partial z_{\beta}^j}]\,\Pi_{\bf z} \right|_{{\bf{z}}=0} \,\, ,
\nonumber
\eq
with $\alpha,\beta=1,\ldots,n$ and $i,j=0,1$.
The relevant projectors are given by $\Pi_{{\bf{z}}}={\cal{U}}({\bf{z}})\Pi\,
{\cal{U}}({\bf{z}})^\dagger,$ where $\Pi$ denotes the projectors over the first $n$
degenerate eigenstates.
Since $\left.{\partial U_\alpha}/
{\partial z^i_{\alpha}}\right|_{z=0}=i^i\,(|\alpha\rangle\langle n+1|+(-1)^i\,|n+1\rangle\langle \alpha|)$,
one finds
\bq
F_{z_\alpha^i z_\beta^j}(0)=i^{i+j}\,[(-1)^j\,|\alpha\rangle\langle\beta|-(-1)^{i}|\beta\rangle\langle\alpha|] \,\, .
\nonumber
\eq
From this expression it follows that
the components of $F$ span the whole $u(n)$ algebra.
As remarked earlier this result does not depend on the specific point chosen, therefore
the case considered is irreducible i.e., $\Gamma_A(L_{\la_0})\cong U(n)$.
Notice that in order to generate the loops in $\cp^n$
one needs to control $2n$ real parameters
instead of the $n^2$ ones necessary for labeling a generic Hamiltonian.

We can now open the way for applying these abstract constructions to quantum computing.
In order to generate a given quantum gate $g\in U(n)$
one has to determine a loop $C_g$ in ${\cal M} \equiv \cp^n$ such that $\Gamma_A(C_g)=g$. 
As the connection of the $\cp^n$ model is irreducible, this is possible 
for any group element $g$.
Due to the non-Abelian character of the connection such an inverse problem
is in general hard to solve. To tackle it we shall take advantage of the 
composition properties described in the previous section.
In particular one chooses specific families of
loops $\{C_i\}$, that generate a complete set of easily 
calculated holonomies, one can eventually construct any $U(n)$
transformation. 
To this end first we consider the $2$-dimensional sub-manifolds in the $2 n$-dimensional
space $(\TH, \PHI)$, spanned by two variables, $(\th_\bi,\phi_{\bar
\bi})$ or $(\th_\bi, \th_{\bar\bi})$, for specific values of $\bi$ and $\bar \bi$. 
For these loops the line integral, given by
\be
\oint _C A = \oint_C ( A^{\th_\bi} d\th_\bi +A^{ \la_{\bar \bi}} d\la_{\bar \bi }) \,\, ,
\nonumber
\ee
where $\la_{\bar \beta} =\th_{\bar \beta}$ or $\phi_{\bar \beta}$, 
includes only two of the $2n$ components of the connection. 
Of course one cannot just simply calculate this line integral and then exponentiate it because
of the path ordering procedure, which is necessary as the matrices
$A^{\th_\bi}$ and $A^{ \la_{\bar \bi}}$ in general do not commute with each other.
To overcome this difficulty we perform a second step for a further restriction of the loops $\{C_i\}$.
From (\ref{niao}) one checks that the parameters which define 
the position of the plane $(\th_\bi , \la_{\bar \bi})$, where the loop $C$ lies, can be always chosen in
such a way that the matrix $A^{\th_\bi}$ is identically zero. 
In particular, if one takes $\th_i=0$, $\forall i\neq \bi, \,\, \bar \bi$, 
the matrices $A^{\th_\bi}$ and $A^{\la_{\bar \bi}}$ commute, so that one can calculate the 
integral and exponentiate it avoiding the path ordering problem.
Of course the choice of the planes has to be
such that the connection components lying on it do not give rise to a trivial holonomy
even if they commute with each other. This is indicated by the 
non-vanishing of the related
field strength component $F_{\th_\bi \la_{\bar \bi}}$. 
Another interpretation of the holonomy, within this approach, is as the exponential 
of the flux of $F$, through the loop $C$. This definition is possible
once the problem has been ``Abelianized'' by having one of the two relevant
components of the connection equal to zero. Application of the Stokes theorem then provides 
 a natural way to evaluate the path integral, as
$\oint_C A^{ \la_{\bar \bi}} d\la_{\bar \bi } = \int_{\D(C)} F_{\th_\bi \la_{\bar \bi}}
d\th_\bi d \la_{\bar \bi}$, where $\D(C)$ is the surface the loop $C$ encloses
on the $(\th_\bi,\la_{\bar \bi})$-plane.

In this framework, it is possible to identify 
four families of loops in such a way as to produce the basis of
four matrices (the Pauli matrices and the identity) of all possible two-by-two 
sub-matrices belonging to the algebra of $U(2)$. With this approach one may 
restrict to a subspace of the degenerate space spanned by the states 
$|\bi \ran$ and $ | \bar \bi\ran$, ordered in such a way that $\bi < \bar \bi$. 
The relevant sets of coordinates are $(\th_\bi ,\, \phi_\bi)$ 
and $(\th_{\bar \bi},\, \phi_{\bar \bi})$.
Taking $\th_i=0$ for all $i\neq \bi,\, \bar \bi$ one obtains the $(\th_\bi, \, \phi_\bi)$
connection components
\end{multicols}
\widetext
\be
\begin{array}{cc}
A^{\th_\bi}=& 
 \left[  \begin{array}{ccc}    0 & 0 \\
                               0 & 0 \\
\end{array} \right] \,\, 
\end{array}
\Sp \text{and} \Sp
\begin{array}{cc}
A^{\phi_\bi}=&  
 \left[  \begin{array}{ccc}    - i \sin ^2 \th_\bi & 0 \\
                               0 & 0 \\
\end{array} \right] \,\, , 
\end{array}
\label{aa1}
\ee
while for the $(\th_{\bar \bi}, \, \phi_{\bar \bi})$ components we have
\be
\begin{array}{cc}
A^{\th_{\bar \bi}}=&  
 \left[  \begin{array}{ccc}    0 & \sin \th_\bi e^{i (\phi_\bi -\phi_{\bar \bi})} \\
                               -\sin \th_\bi e^{-i (\phi_\bi -\phi_{\bar \bi})} & 0 \\
\end{array} \right]  
\end{array}
\label{aa2}
\ee
and
\be
\begin{array}{cc}
A^{\phi_{\bar \bi}}=&  
 \left[  \begin{array}{ccc}    i \sin ^2 \th_\bi \sin ^2 \th_{\bar \bi} & -i \sin \th_\bi \sin \th_{\bar \bi} \cos \th_{\bar \bi} e^{i(\phi_\bi -\phi_{\bar \bi})} \\
                               -i \sin \th_\bi \sin \th_{\bar \bi} \cos \th_{\bar \bi} e^{-i(\phi_\bi -\phi_{\bar \bi})} & -i \sin ^2 \th_{\bar \bi}\\
\end{array} \right] \,\, .
\end{array}
\label{aa3}
\ee
\begin{multicols}{2}
\narrowtext
With these four matrices we want to build a complete set of generators for the $U(2)$ group. 
Specific choices of coordinate planes inside the four dimensional sub-manifold with coordinates
$\{\th_\bi , \, \phi_\bi, \, \th_{\bar \bi} ,\, \phi_{\bar \bi}\}$ shall provide 
those matrices.
Note that this sub-manifold is locally isomorphic to $\cp^2,$ and due to the irreducibility
of its relevant connection it is possible to produce the whole $U(2)$ group.

We can calculate the components of the field strength associated with the
connection components (\ref{aa1}),(\ref{aa2}), (\ref{aa3}). With a straightforward application 
of (\ref{curvature}) we obtain
the non-zero field strength components to be
\end{multicols}
\widetext
\bq
&&
\begin{array}{cc}
F_{\th_\beta \phi_\beta}=-i&
 \left[  \begin{array}{ccc}    \sin 2 \th_\beta & 0 \\
                               0 & 0 \\
\end{array} \right]
\end{array}
\,\,\, , \Sp
\begin{array}{cc}
F_{\th_\beta \phi_{\bar \bi}}=i&  
 \left[  \begin{array}{ccc}    \sin 2 \th_\bi \sin ^2 \th_{\bar \bi} &
                 -{1 \over 2}\cos \th_\bi \sin 2 \th_{\bar \bi} e^{i(\phi_\bi -\phi_{\bar \bi})}\\
                     -{1 \over 2}\cos \th_\bi \sin 2 \th_{\bar \bi} e^{-i(\phi_\bi -\phi_{\bar \bi})} 
                                        & 0 \\
\end{array} \right] \,\, , 
\end{array}
\nonumber
\eq
\bq
&&
\begin{array}{cc}
F_{\th_\beta \th_{\bar \bi}}=\cos \th_\bi &
 \left[  \begin{array}{ccc}    0 & e^{i(\phi_\bi -\phi_{\bar \bi})} \\
                               -e^{-i(\phi_\bi -\phi_{\bar \bi})} & 0 \\
\end{array} \right] 
\end{array}
\,\,\, , \Sp
\begin{array}{cc}
F_{\th_{\bar \bi} \phi_\bi}=-i\sin \th_\bi \cos^2 \th_\bi &  
 \left[  \begin{array}{ccc}    0 & e^{i(\phi_\bi -\phi_{\bar \bi})}\\
                     e^{-i(\phi_\bi -\phi_{\bar \bi})} & 0 \\
\end{array} \right] \,\, , 
\end{array}
\nonumber
\eq
\bq
&&
\begin{array}{cc}
F_{\th_{\bar \bi} \phi_{\bar \bi}}=i &
 \left[  \begin{array}{ccc}    0 & 
                \sin \th_\bi \cos^2 \th_\bi \sin^2 \th_{\bar \bi} e^{i(\phi_\bi -\phi_{\bar \bi})} \\
        \sin \th_\bi \cos^2 \th_\bi \sin^2 \th_{\bar \bi} e^{-i(\phi_\bi -\phi_{\bar \bi})} & 
                                        -\cos^2 \th_\bi \sin 2\th_{\bar \bi} \\
\end{array} \right] 
\end{array}
\nonumber
\eq
and
\bq
&&
\begin{array}{cc}
F_{\phi_\bi \phi_{\bar \bi}}= {1 \over 2} \sin \th_\bi \cos^2 \th_\bi \sin 2 \th_{\bar \bi} &  
 \left[  \begin{array}{ccc}    0 & e^{i(\phi_\bi -\phi_{\bar \bi})}\\
                     -e^{-i(\phi_\bi -\phi_{\bar \bi})} & 0 \\
\end{array} \right] \,\, .
\end{array}
\nonumber
\eq
\begin{multicols}{2}
\narrowtext
The field strength will be used in the following  section to calculate the holonomies.

\subsection{The Holonomies $\Gamma_A(\gamma)$}

Let us see how the restriction on various planes affects the relevant connection 
components (\ref{aa1}),(\ref{aa2}),(\ref{aa3}).
It is logical to choose one of the plane coordinates to be the $\th_\bi$ one as 
$A_{\th_\bi}\equiv {\bf 0}$ and hence it commutes with all others   components,
as required for implementing our strategy. 
The first choice is the plane $(\th_\bi, \phi_\bi)$, where
the non-zero component of the connection is $A^{\phi_\bi}_{\bi \bi}=-i \sin^2 \th_\bi$.
The second choice is the plane $(\th_\bi, \phi_{\bar \bi})$ for $\bar \bi > \bi$, 
with $\th_{\bar \bi}=\pi/2,$ giving a
different connection with two non-zero elements, 
$A^{\phi_{\bar \bi}} _{\bi \bi}=i \sin^2 \th_\bi$ and 
$A^{\phi_{\bar \bi}} _{\bar \bi \bar \bi}=-i$. Of course the latter element will give zero
when integrated along a loop. 
For $\bi > \bar \bi$ both matrices are
identically zero, and give rise to a trivial holonomy. With these two components
and for appropriate loops one can
obtain all possible $U(n)$ {\it diagonal} transformations. Indeed, for the loop
$C_1 \in (\th_\bi, \phi_\bi)$ we obtain
\bq
\Gamma_A(C_1)= \exp [ -i |\bi \ran \lan \bi | \Sigma_1 ]
\nonumber
\eq
 $\Sigma_1$ denoting the
area enclosed by $C_1$ on a $S^2$ sphere with coordinates $(2\th_\bi,\phi_\bi)$. 
This is exactly the Abelian Berry phase which could be produced if the state $|\bi\ran$
were non-degenerate as it does not get mixed with the rest of the states. For
$C_2 \in (\th_\bi, \phi_{\bar \bi})$ we obtain analogously the holonomy
\bq
\Gamma_A(C_2)= \exp [i |\bi \ran \lan  \bi | \Sigma_2] \,\, .
\nonumber
\eq
Recalling the constraint $ \bi< \bar \bi$, we see that one can produce $n-1$ distinct holonomies
from $C_2$ type  loops. 

To obtain the non-diagonal transformations one has to consider a loop on the $(\th_\bi,
\th_{\bar \bi})$ plane, with $\th_i=0$ for all $i\neq \bi,\bar \bi$. Then, the only non-vanishing
elements of the connection are $A^{\th_{\bar \bi}}_{\bi \bar \bi}=e^{i(\phi_{\bi}-\phi_
{\bar \bi})} \sin \th_\bi= -\bar A^{\th_{\bar \bi}}_{\bar \bi \bi}.$
By choosing further the 
$(\theta_\beta,\,\th_{\bar \bi})$ plane at the position $\phi_\bi=\phi_{\bar \bi}=0$ the 
holonomy becomes, for the loop $C_3 \in (\th_\bi, \th_{\bar \bi})_{\phi_\bi=\phi_{\bar \bi}=0}$
\bq
\Gamma_A(C_3)= \exp [ -i(-i|\bi \ran \lan \bar \bi| + i|\bar \bi \ran \lan \bi |)\tilde \Sigma_3 ]\,\, ,
\nonumber
\eq
while at $\phi_\bi=\pi /2$ and $\phi_{\bar \bi}=0$, i.e. 
$C_4 \in (\th_\bi, \th_{\bar \bi})_{\phi_\bi=\pi/2,\phi_{\bar \bi}=0}$ we have 
\bq
\Gamma_A(C_4)= \exp [-i (| \bi \ran \lan \bar \bi| + |\bar \bi \ran \lan \bi |) \tilde \Sigma _4 ] \,\, ,
\nonumber
\eq
where $\tilde\Sigma$ is the area on the sphere with coordinates 
$(\pi/2 -\th_\bi, \th_{\bar \bi})$. The positive or negative sign in front 
of the area depends on the orientation of the surface enclosed by the loop $C$ 
with respect to the orientation of the field strength $F$.
Note that any loop $C$ on the $(\th_\bi, \la_{\bar \bi})$
plane with the same enclosed area $\Sigma_C$ (when mapped on the appropriate sphere) 
will give the same holonomy independent of its position and shape. 
These four holonomies restricted each time to a specific $2\times2$ sub-matrix 
generate all $U(2)$ transformations. 
Hence, considering the inverse problem of obtaining a desired unitary from a holonomy
we are able to choose from a whole family of loops of arbitrary shape and position.

Finally, it is easy to check that in this way one can indeed obtain
$U=\exp[\mu_a\, T_a]$, where $T_a\,(a=1,\ldots,n^2)$ is a $u(n)$ anti-hermitian generator
and $\mu_a$ an arbitrary real number.
Therefore any element of $U(n)$ can be obtained
by controlling the $2n$ parameters labeling the points of ${\bf{CP}}^n.$

It is instructive to consider the form that the Hamiltonian family $\cal
F$ takes when restricted to the particular 2-sub-manifolds. 
For the loop $C_1$ (and similarly for $C_2$) one finds 
\bq
H_1=-\varepsilon/2\, \vec{B}(2\th_{\bi},\phi_{\bi})\cdot\vec{\hat \sigma}
\nonumber
\eq
for $\vec{B}(\th_i,\phi_j) =(\sin \th_i \cos \phi_j, \sin \th_i \sin \phi_j,
\cos \th_i)^T$, where the only non-zero elements are on the
$\bi$-th and $(n+1)$-th row and column. $H_1$ generates an Abelian ${\bf CP}^1$
phase in front of the state $|\bi\ran$ and the conjugate one in front of $|n+1 \ran$. 
On the other hand for the path $C_3$ (and similarly for $C_4$) we have 
\bq
H_3=\varepsilon {\vec B}(\th_\bi,\th_{\bar \bi}) {\vec B}(\th_\bi,\th_{\bar \bi})^T \,\, ,
\nonumber
\eq
where the non-zero elements connect the states $|\bi\ran$, $|\bar \bi \ran$
and $|n+1\ran$. In this 
Hamiltonian there is direct coupling between three states, giving rise to a
non-Abelian interaction. While $H_1$ is easy to simulate in the laboratory with 
various experimental setups (spin $1/2$ particle in a magnetic field, NMR, 
optical polarization, etc) it is  yet an open  challenge to construct the interaction 
dictated by the Hamiltonian $H_3$.

\section{Application to the Optical Holonomic Setup}

In the previous sections we have developed the theoretical background in order
to make  the non-Abelian geometrical phases useful into the quantum computing arena.
Complex as it may be, such a construction offers various possibilities 
and advantages when applied to physical systems. In particular we shall
resorts to  quantum optics in order to make a physical application of HQC \cite{hol3},
but also to clarify and resolve some theoretical issues discussed in the
previous sections. To this aim  we employ   existing devices of quantum optics, 
such as displacing and squeezing devices and interferometers, acting 
on laser beams in a non-linear medium.

\subsection{Displacers, Squeezers and Interferometers as holonomic devices \label{comm}}

In this subsection we shall consider the realistic implementation of the 
holonomic computation in the frame of  quantum optics. Even though the complete 
implementation of the model presented here is likely to be experimentally a very challenging task,
it is still remarkable that the necessary employed devices are 
realizable in the laboratory. Let us briefly present the setup.

In order to perform holonomic computation with laser beams we shall consider 
the non-linear interaction Hamiltonian produced by a Kerr medium
\bq
H_I=\hbar X n(n-1) \,\, ,
\nonumber 
\eq
with $n=a^\dagger a$ the number operator, $a$ and $a^{\dagger}$ being the usual 
bosonic annihilation and creation operators respectively, and $X$ a constant 
proportional to the third order nonlinear susceptibility, $\chi^{(3)}$, of the medium. 
The degenerate eigenstates of $H_I$ are the $|0\ran$ and $|1\ran$, where 
$\{|\nu\ran ; \nu=0,1,... \}$ denote the Fock basis of number eigenstates, 
$n|\nu\ran=\nu |\nu \ran$. The degenerate space they span will be the 
one qubit coding space. For the tensor product structure of $l$ qubits we
have to employ a set of $l$ beams, providing the
basis states $|\nu_1...\nu_m\ran=|\nu_1\ran \otimes...\otimes |\nu_m\ran$ 
where $\nu_l$ could be zero or one, for $l=1,...,m$. The code can be written in this space of states. 

The iso-spectral transformations of the Hamiltonian, $H(\sigma)=\U(\si) H_I\U^\dagger(\si)$, can be constructed 
by resorting to displacing and squeezing devices with unitaries 
$D(\la)=\exp(\la a^\dagger-\bar \la a)$ and $S(\mu)=\exp(\mu {a^\dagger}^2-\bar \mu a^2)$
respectively, as well as two mode displacing and squeezing devices:
$N(\xi)=\exp(\xi a_1^\dagger a_2 -\bar \xi a_1 a_2^\dagger)$ and 
$M(\zeta)=\exp(\zeta a_1^\dagger a_2^\dagger -\bar \zeta a_1 a_2)$ respectively.
In fact to obtain holonomies for one qubit gates we employ
$\U(\si)=D(\la)S(\mu)$ while for two qubits we take $\U(\si)=N(\xi) M(\zeta)$.
In the first case the connection components are two-by-two matrices,
where the null action is assumed on the rest of the tensor product sub-systems.
For the coordinate decomposition $\la=x+iy$ and $\mu=r_1 \exp i \th_1$ we have
after some simple algebra, similar to the one for $\cp^2$, the connection components
\end{multicols}
\widetext
\[ \begin{array}{cc}
A_x=&  
 \left[  \begin{array}{ccc}     -iy & -(\cosh 2r_1 - e^{i\th_1} \sinh 2r_1) \\
                                \cosh 2r_1-e^{-i\th_1} \sinh 2r_1& -iy \\
\end{array} \right] \,\, ,
\end{array}\]

\[ \begin{array}{cc}
A_y=&  
 \left[  \begin{array}{ccc}     ix & i(\cosh 2r_1 + e^{i\th_1} \sinh 2r_1) \\
                                i(\cosh 2r_1 + e^{-i\th_1} \sinh 2r_1) & ix \\
\end{array} \right] \,\, ,
\end{array}\]

\[ \begin{array}{ccc}
A_{r_1}=&  
 \left[  \begin{array}{ccc}    0 & 0 \\
                               0 & 0 \\
\end{array} \right] \,\,\,\,\,\, ,\,\,\,\,\,\,\,\,\,\,\,\,\,\,
A_{\th_1}=&  
 \left[  \begin{array}{ccc}    1 & 0 \\
                               0 & 3 \\
\end{array} \right] {i \over 4} (\cosh 4 r_1 -1)  \,\, .
\end{array}\]
\begin{multicols}{2}
\narrowtext
The diagonal elements of the above matrices are also given easily by the equivalent
Berry phases acquired by non-degenerate Fock states $|\nu\ran$. For example the 
Abelian phases produced by a displacer are equal for all Fock states and given by \cite{indians}
\be
\phi_{Berry}^\nu = \oint (ydx-xdy) \,\, ,
\label{disp}
\ee
while the phases produced by a squeezer are given by
\bq
\phi_{Berry}^\nu  = \frac{2\nu+1}{4} \oint \left( \cosh 4r_1 -1 \right)d \th_1 \,\, .
\nonumber
\eq
Notice the complete similarity of the generators of the squeezing phases with
$A_{\th_1}$ as the non-Abelian matrix is diagonal, while in the displacing case
only the diagonal elements are reproduced by (\ref{disp}).

On the other hand the parametric space of the
two mode interferometers gives rise
to the following connection components
\[ \begin{array}{ccc}
A_{r_2}=&  
 \left[  \begin{array}{cccc}    0 & 0 & 0 & -e^{-i\th_2}\\
                                0 & 0 & 0 & 0 \\
                                0 & 0 & 0 & 0 \\
                                e^{i\th_2} & 0 & 0 & 0 \\    
\end{array} \right] \,\, ,
\end{array}\]
\[ \begin{array}{ccc}
A_{r_3}=&  
 \left[  \begin{array}{cccc}    0 & 0 & 0 & 0 \\
                                0 & 0 & -e^{-i\th_3} & 0 \\
                                0 & e^{i\th_3} & 0 & 0 \\
                                0 & 0 & 0 & 0 \\    
\end{array} \right] (2 \cosh ^2 r_2 -1) \,\, .
\end{array}\]
as well as
\[ \begin{array}{cccc}
A_{\th_2}=&  
\left[  \begin{array}{cccc}    0 & 0 & 0 & e^{-i\th_2}\\
                                0 & 0 & 0 & 0 \\
                                0 & 0 & 0 & 0 \\
                                e^{i\th_2} & 0 & 0 & 0 \\    
\end{array} \right] {i \over 2} \sinh 2 r_2 \Sp + \Sp 
\end{array}\]
\[ \begin{array}{ccc}
\Sp \left[  \begin{array}{cccc}     1 & 0 & 0 & 0 \\
                                0 & 2 & 0 & 0 \\
                                0 & 0 & 2 & 0 \\
                                0 & 0 & 0 & 3 \\    
\end{array} \right] {i \over 2} ( \cosh 2 r_2 -1) \,\, ,
\end{array}\]
and
\[ \begin{array}{cccc}
A_{\th_3}=&
\left[  \begin{array}{cccc}    0 & 0 & 0 & 0 \\
                                0 & 0 & e^{-i\th_3} & 0 \\
                                0 & e^{i\th_3} & 0 & 0 \\
                                0 & 0 & 0 & 0 \\    
\end{array} \right] {i \over 2} \cosh 2 r_2 \sin 2 r_3 \,\,\,\, + \Sp
\end{array}\]
\[ \begin{array}{ccc}
\Sp \left[  \begin{array}{cccc}     0 & 0 & 0 & 0 \\
                                0 & 1 & 0 & 0 \\
                                0 & 0 & -1 & 0 \\
                                0 & 0 & 0 & 0 \\    
\end{array} \right] i \sin^2 r_3 \,\, ,
\end{array}\]
where $\zeta=r_2 \exp i \th_2$ and $\xi=r_3 \exp i \th_3$. 
Here we define ${\cal M} := \{ \sigma_i\}$ th $\sigma_i$'s being the real
coordinates $\{x,y,r_i, \theta_i\}$ parameterizing all possible configurations with devices 
acting on the various laser beams.
The first two components
are enough for constructing holonomies representing two qubit gates between any two qubits,
which together with the one qubit rotations 
result into a universal set of gates. It is straightforward to apply
the conditions posed in the previous section for finding coordinate planes, with 
the relevant connection components commuting. As $A_{r_1}$ is identically zero 
it can be combined with $A_x$, $A_y$ and $A_{\th_1}$ to produce $U(2)$ 
holonomies, while $A_{r_2}$ and $A_{r_3}$ commute with each other leading to generate
$U(4)$ matrices. 

The holonomies obtained from the various loops are given in the following.
The loop $C_I \in \left. (x,r_1)\right._{\th_1=0}$ gives
\bq
\Gamma_A(C_I)=\exp -i\hat \sigma_1 \Sigma_I 
\nonumber
\eq
with $\Sigma_I:=\int_{D(C_I)} \!dxdr_1 2 e^{-2r_1}$.
The loop $C_{II} \in \left. (y,r_1)\right._{\th_1=0}$ gives
\bq
\Gamma_A(C_{II})=\exp -i\hat \sigma_2 \Sigma_{II} \,\, ,
\nonumber
\eq
with area $\Sigma_{II}:=\int_{D(C_{II})} \! dydr_1 2 e^{2r_1}$.
The loop $C_{III} \in (r_1,\th_1)$ gives 
\bq
\Gamma_A(C_{III})=\exp -i\hat{s}_3 \Sigma_{III} \,\, ,
\nonumber
\eq
with $\Sigma_{III}:=\int_{D(C_{III})} \! dr_1d\th_1 \sinh 4r_1$.
$C_I$, $C_{II}$ and $C_{III}$ produce $U(2)$ rotations. 
In order to generate $U(4)$ group elements we span the following configurations.
The loop $C_{IV} \in \left. (r_2,r_3)\right._{\th_2=\th_3=0}$ gives
\bq
\Gamma_A(C_{IV})=\exp -i\hat \sigma_2^{12} \Sigma_{IV}
\nonumber
\eq
with $\Sigma_{IV}:=\int_{D(C_{IV})} \! dr_2dr_3 2 \sinh 2 r_2$.
The loop $C_{V} \in \left. (r_2,r_3)\right._{\th_2=0, \th_3=3\pi/2}$
gives
\bq
\Gamma_A(C_{V})=\exp -i\hat \sigma_1^{12} \Sigma_{V}
\nonumber
\eq
with the area given by $\Sigma_{V}:=\int_{D(C_{V})} \! dr_2dr_3 2 \sinh 2 r_2$.
Above we have used
\bq
\hat{s}_3:= - \left[  \begin{array}{cccc}    1 & 0 \\
                                                0 & 3 \\
\end{array} \right]
\Sp , \Sp
\hat \sigma _2^{12}:=
\left[  \begin{array}{cccc}    0 & 0 & 0 & 0 \\
                                0 & 0 & -i & 0 \\
                                0 & i & 0 & 0 \\
                                0 & 0 & 0 & 0 \\
\end{array} \right]
\nonumber
\eq
and
\bq
\hat \sigma _1^{12}:=
\left[  \begin{array}{cccc}    0 & 0 & 0 & 0 \\
                                0 & 0 & 1 & 0 \\
                                0 & 1 & 0 & 0 \\
                                0 & 0 & 0 & 0 \\
\end{array} \right] \,\, ,
\nonumber
\eq
while $D(C_\rho)$ with $\rho=I,...,V$ is the surface on the relevant 
sub-manifold $(\sigma_i,\sigma_j)$ of ${\cal M}$ whose boundary is the path $C_\rho$. 
The hyperbolic functions in these integrals stem out of the geometry of the $su(1,1)$ 
manifold associated with the relevant control sub-manifold. The $\Gamma_A(C)$'s thus 
generated belong either to the $U(2)$ or $U(4)$ group and act on the one qubit space 
or on the space of the tensor product of two qubits, respectively. Considering the 
tensor product structure of our system these rotations represent in the $2^m$ space 
of $m$ qubits respectively single qubit rotations and two qubit interactions, thus 
resulting into a universal set of logical gates. For example, $\Gamma_A(C_{V})$ with 
$\Sigma_V=\pi/4$ gives the following non-trivial two qubits gate
\bq
U= {1 \over \sqrt{2}}
\left[  \begin{array}{cccc}     \sqrt{2} & 0 & 0 & 0 \\
                                0 & 1 & -i\,\, & 0 \\
                                0 & -i\,\, & 1 & 0 \\
                                0 & 0 & 0 & \sqrt{2} \\
\end{array} \right]
\nonumber
\eq
while the holonomies produced by the loops $C_I$, $C_{II}$ 
and $C_{III}$ give a general one qubit rotation. Together they compose a universal 
set of transformations \cite{Loss}. 

The above description can be applied also to quantum systems with harmonic
oscillator structure. For example, ion traps or atoms in a cavity are
described by oscillating modes (the ionic vibration or the cavity mode), 
which can be also displaced, squeezed and
interfered with standard experimental techniques
\cite{Meekhof,Cirac,Steinbach,Alsing}. The degeneracy can be provided by the
coupling of the oscillating modes with the internal energy levels of the atoms
or by the structure of the product space of states of two modes \cite{JP}.

\subsection{$SU(2)$ Interferometers and non-Abelian Stokes Theorem}

In what follows we discuss the possible use  of $SU(2)$ interferometer as a control 
 devices \cite{Yurke} for producing holonomies. 
While relevant to quantum optical devices, such a scheme has an additional 
theoretical interest. Rather than distilling the Abelian sub-structure to avoid 
the path ordering problem we shall employ the non-Abelian Stokes theorem in order
to rewrite the loop integral of the connection as a surface integral of its field 
strength. The main advantage of this approach is the partial relaxation of the path 
ordering conditions, enabling the holonomic calculation of non-commuting connection
components.

For $a_1$ and $a_2$ the annihilation 
operator of two different laser beams, consider the Hermitian operators
\bq
&
J_x={1 \over 2} (a_1^{\dagger} a_2 +a_2^\dagger a_1)
\,\, , \,\,\,
J_y=-{i \over 2} (a_1^{\dagger} a_2 -a_2^\dagger a_1)
\,\, , &
\no \no
&
J_z={1 \over 2} (a_1^{\dagger} a_1 -a_2^\dagger a_2)
&
\label{JJJ}
\eq
and
\bq
N=a_1^{\dagger} a_1 +a_2^\dagger a_2=n_1+n_2 \,\, .
\nonumber
\eq
The operators (\ref{JJJ}) satisfy the commutation relations for the Lie 
algebra of $SU(2)$; $[J_x,J_y]=iJ_z$, $[J_y,J_z]=iJ_x$, $[J_z,J_x]=iJ_y$. 
The operator $N$, which is proportional to the free Hamiltonian of two 
laser beams, commutes with all of the $J$'s. On the other hand, however, 
the Kerr Hamiltonian does not commute with the $J$'s, allowing for the 
possibility that $SU(2)$ interferometers be used as transformation 
controllers in view of the holonomic computation. 

From the above operators we obtain the unitaries, $U_x(\al)=\exp(i\al J_x)$, 
$U_y(\beta)= \exp (i \beta J_y)$ and $U_z(\gamma)=\exp(i\gamma J_z)$. 
For the degenerate state space of two laser beams spanned by $|\nu_1 \nu_2\ran$, 
we have from (\ref{conn}) and for ${\cal U}=U_x(\al)U_y(\beta)U_z(\gamma)$ 
the following connection components
\bq
A_\al ={i \over 2}
 \left[  \begin{array}{cccc}    0 & 0 & 0 & 0 \\
                                0 & \sin \beta & \cos \beta e^{i\gamma} & 0 \\
                                0 & \cos \beta e^{-i \gamma} & -\sin \beta & 0 \\
                                0 & 0 & 0 & 0 \\
\end{array} \right] \,\, ,
\nonumber
\eq
\bq
A_\beta=-{1 \over 2}
 \left[  \begin{array}{cccc}    0 & 0 & 0 & 0 \\
                                0 & 0 & e^{i\gamma} & 0 \\
                                0 & -e^{-i \gamma} & 0 & 0 \\
                                0 & 0 & 0 & 0 \\
\end{array} \right] 
\nonumber
\eq
and
\bq
A_\gamma=-{i \over 2}
 \left[  \begin{array}{cccc}    0 & 0 & 0 & 0 \\
                                0 & 1 & 0 & 0 \\
                                0 & 0 & -1 & 0 \\
                                0 & 0 & 0 & 0 \\
\end{array} \right] .
\nonumber
\eq
These components do not commute with each other, when projected on planes 
with non-trivial field strength. Hence, it is not possible to adopt again 
the method used previously to calculate the holonomies of 
paths in the three dimensional parameter space, $(\al, \beta,\gamma)$. 
Instead, for this purpose we can employ the non-Abelian Stokes theorem \cite{Karp}. 
The extra limitation, now, for the choice of the path comes from the constraint 
that, apart from being confined on a special two dimensional subspace, it has 
to have the shape of an orthogonal parallelogram with two sides lying along 
the coordinate axis. This will facilitate the extraction of an analytic result 
from the Stokes theorem. With the loop composition properties
described earlier, it is possible to generalize these simple structures to 
any desirable loop. 

In order to state the non-Abelian Stokes theorem let us first present some preliminaries, 
where a few simplifications are introduced, as its general form will not be 
necessary in the present work. 
Consider the Wilson loop (holonomy), $W={\bf P}\exp \oint_C A$, of a rectangular
loop $C$ with the sides parallel to the coordinates $(\sigma, \tau)$
where $\sigma$ and $\tau$ are a parameterization of the plane where the
loop $C$ lies. $W$ is made out of
the Wilson lines $W_i$ for $i=1,...,4$, as $W=W_4 W_3 W_2
W_1$, where $W_i$ corresponds to the $i$'th side of the rectangular
ordered in anti-clockwise way. Define $T^{-1}(\sigma,\tau) \equiv W_4 W_3$. Then, for
$F_{\sigma\tau}$ the field strength of the connection $A$ on the plane
$(\sigma,\tau)$, $W$ is given in terms of a surface integral
\bq
&&
W={\bf P}_\tau e^{\int_\Sigma T^{-1}(\sigma,\tau)F_{\sigma \tau}
(\sigma,\tau)T(\sigma,\tau) d\sigma d\tau} \,\, ,
\label{stokes}
\eq
where ${\bf P}_\tau$ is the path ordering symbol with respect only to
the $\tau$ variable contrary to the usual path ordering symbol {\bf
P} which is defined with respect to both variables, $\sigma$ and $\tau$. Here
$F_{\sigma \tau}(\sigma,\tau)=-\p_\sigma A_\tau +\p_\tau
A_\sigma+[A_\sigma , A_\tau]$.

The key point of this approach is that in (\ref{stokes}) the path ordering is only with 
respect to $\tau$ so there is the possibility to arrange the exponent in such a way that its 
integration with respect to $\tau$ is performed with a trivial path ordering procedure. 
Such possibility can be achieved in the present case by choosing the integration limits with respect to 
$\si$ so that they give commuting matrices as functions of $\tau$. 
In detail, we first choose a loop $C_1$ on the plane $(\al,\beta)$ positioned at $\gamma=0$. 
The relevant field strength component and the matrix $T^{-1}(\al, \, \bi)$ are given by
\bq
F_{\al \bi}=i (\cos \bi \hat \si_3 -\sin \bi \hat \si_1)
\nonumber
\eq
and 
\bq
T^{-1}= \exp({i {\bi \over 2} \hat \si_2} )
\exp\left[{-i {\al \over 2} (\sin \bi \hat \si_3 +\cos \bi \hat \si_1)}\right]   \,\, .
\nonumber
\eq
Hence we obtain
\bq
&&
T^{-1} F_{\al \bi} T =
\no \no
&&
-i (\cos \al \sin 2 \bi \, \hat \si_1 -\cos \al \cos 2 \bi \, 
\hat \si_3 + \sin \al \, \hat \si_2 )
\eq
By taking a rectangle with the $\al$-side equal to $\pi$ the only remaining term 
after the $\al$ integration is the one with $\hat \si_2$, and hence the path ordering with respect 
to $\bi$ does not pose any computational  problem. As a result we have that for the closed rectangular 
loop $C_1\in (\al,\beta)$-plane with coordinates $\{(0,0),(\al \! = \! \pi,0),
(\al \! = \! \pi,\beta),(0,\beta) \}$ we obtain the following unitary transformation
\be
\Gamma_A(C_1)=\exp (-i2\beta \hat \sigma_2^{12}) \,\, .
\label{C1}
\ee
With similar reasoning a rectangular loop $C_2\in (\al,\gamma)$-plane 
with coordinates $\{(0,0),(\al \! = \! \pi,0),(\al \! = \! \pi,\gamma),(0,\gamma)\}$ 
has the field strength component and the matrix $T^{-1}(\al, \, \gamma)$ given by
\bq
F_{\al \gamma}=-i (\cos \gamma \hat \si_2 +\sin \gamma \hat \si_1)
\nonumber
\eq
and
\bq
T^{-1}= \exp({i {\gamma \over 2} \hat \si_3} )
\exp\left[{-i {\al \over 2} (\cos \gamma \hat \si_1 - \sin \gamma \hat \si_2)}\right]   \,\, ,
\nonumber
\eq
which give finally the holonomy
\bq
\Gamma_A(C_2)=\exp (-i2\gamma \hat \sigma_3^{12}) \,\, .
\nonumber
\eq
The matrix $\hat \sigma_3^{12}$ is defined similarly to $\hat \sigma_1^{12}$ and 
$\hat \sigma_2^{12}$ in Subsection \ref{comm}. Note that the coefficients in front 
of the matrices in the unitaries are areas on spheres spanned by the angles $\al$ 
and $\beta$ or $\al$ and $\gamma$. This is consistent with the geometry of $SU(2)$. 

Due to their structure the two matrices $\Gamma_A(C_1)$ and $\Gamma_A(C_2)$ can produce 
any unitary transformation of {\it one} qubit 
encoded in the sub-space of states of two laser beams spanned by $\{|01\ran,|10\ran\}$. 
In other words, we need two laser beams to encode one qubit, contrary to previous 
construction. The operation of two qubit interactions demands sophisticated optical 
devices as four mode interferometers, and we shall not discuss them here. 

\subsection{Holonomies and Devices}

In the optical example studied above, a general state 
$|\psi\ran$ in the degenerate eigen-space of $H_0=H_{Kerr}$ is given as a linear 
combination of $|0\ran$ and $|1\ran$. 
The interaction Hamiltonian $H_{Kerr}$ describes fully our system when we move
to the interaction picture. This transition affects only the parameters of the
control devices which are redefined. Under an iso-spectral, cyclic and adiabatic evolution of 
the Hamiltonian in the family ${\cal F}$, the evolution operator acting on 
$|\psi\ran$ is given by the $2 \times 2$ sub-matrix in the upper left corner of the matrix
\bq
U(0,T)={\bf T} \exp -i \int ^T_0 {\cal U}(\sigma (t))H_0 {\cal U}^\dagger(\sigma(t))dt \,\, .
\label{steps}
\eq
This evolution takes place from time $0$ to time $T$ and, to perform a closed path $\gamma$, 
we demand $\sigma(0)=\sigma(T)$. 
The iso-spectral rotations of the Hamiltonian can be achieved with the use of the optical 
control devices described in the previous subsections.
The degeneracy is provided by the propagation of the laser beams
through the Kerr medium. Hence, when the laser beams are in the control
devices, the degeneracy is not present. To restore degeneracy along the adiabatic
evolution one may employ the kicking method \cite{hol2}, \cite{SYMM}
 now briefly recalled.

Suppose that one is able
to turn on and off a set of interactions (the ``kicks'')
$K:=\{{\cal U}(\sigma)\}_{\sigma\in{\cal M}}$
on an ultra-fast time-scale with respect to the unperturbed part of the evolution. 
Let $T=N\,\Delta\,t$ and $t_0=0,\,t_{i+1}=t_i+\Delta\,t\,(i=1,\ldots\,N-1)$
be a partition of the time interval $[0,\,T].$
The system evolution is as follows:
at any time $t_i$ one kicks the system with the pulse
${\cal U}^\dagger_{i+1}\,{\cal U}_{i}\,$ where ${\cal U}_i:={\cal U}(\sigma_i)$ is a
unitary belonging to $K$ (${\cal U}_0={\cal U}_N=\openone$).
If between the kicks the system evolution is unperturbed,
one obtains the evolution along the loop $\gamma$ by
\bq
U_\gamma
%U_{N,\Delta t} (T)
={\bf {T}}\,\prod_{i=1}^{N-1}{\cal U}_i\,e^{-iH_0 \Delta t} 
\,{\cal U}_i^\dagger ,
\nonumber
\eq
In the limit $\Delta t\mapsto 0,\,N\mapsto\infty$,
[$N\,\Delta t=T$]
one gets
$ U_{\gamma} \rightarrow {\bf {T}} \exp\,\int_0^T dt\, H(t)$
where $H(t):= {\cal U}(\sigma(t))\,H_0\,{\cal U}(\sigma(t))^\dagger.$
By making the function $\sigma(t)$ vary adiabatically,
one can approximate the holonomic evolution.
The latter simulates 
the desired evolution (\ref{steps}) by alternating the action of the 
degenerate Hamiltonian, $H_0$, and the control procedure with infinitesimal steps.
We shall employ it here to demonstrate in a simple example the fidelity of 
the evolution operator approximated by the kicking method procedure
and the relevant holonomy theoretically predicted.

Let us consider displacing 
devices $D(\la)$, performing a closed loop in their control parameters $\la$. 
\begin{figure}[hb]
 \epsffile{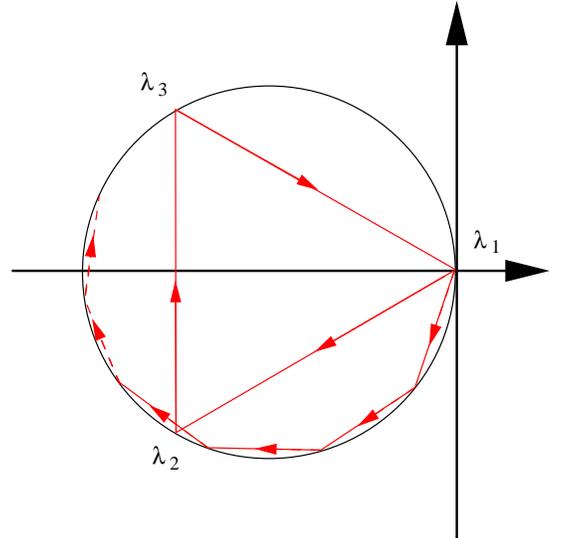}
  \caption[contour]{\label{trig}
The triangular (and polygonal) loop $C$ on the complex plane of the displacing 
control parameters, $\la$, approximating the loop (circle).
           }
\end{figure}
This is shown in Fig. \ref{trig}, where for simplicity the least possible 
number of displacing devices (three) for performing a closed loop has been 
drawn. When the state is at the edges of the polygon (triangle) 
no displacement takes place and the action
of the Kerr medium is implied. Two displacing unitaries are combined as 
$D(\lambda)D(\lambda')=\exp{(i \Im (\lambda \bar \lambda'))} D(\la+\la')$. 
The physical process behind this is as follows. On the state $|\psi\ran$ 
first acts a displacing unitary $D^\dagger(\la_1)$, taking it 
to the point $\la_1$. Then, the evolution operator of the Kerr Hamiltonian, 
$U(\Delta t)=\exp( -i H_0 \Delta t)$,
acts for a time interval $\Delta t= T/3$. 
This effect is achieved by propagating the beam inside a Kerr medium. Then, 
the evolution $D^{\dagger}(\la_2)D(\la_1)$ is performed. This is obtained, 
with a single displacing device, given (up to an overall phase factor that 
will cancel at the end) by $D(\la_1-\la_2)$. After exiting the displacing 
device (we are at point $\la_2$) the beam enters a Kerr medium once more for time 
$\Delta t$ and then the procedure is repeated until one comes back to the point 
$\la_1$ and the beam enters again the Kerr medium. Eventually, the state is 
thus displaced by $D(\la_1)$. This loop may be transported to any other place 
of the control parameter complex plane by acting at the beginning and at the 
end of this procedure with the appropriate displacing unitary (device).
\end{multicols}
\widetext
\begin{figure}[hb]
  \epsffile{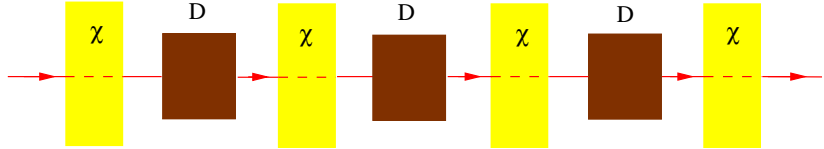}
  \caption[contour]{\label{series}
The setup with the alternating Kerr media and Displacing devices.
           }
\end{figure}
\begin{multicols}{2}
\narrowtext
The evolution operator is approximated by the following evolution operator
\bq
&&
U(0,T)\approx D(\la_1) \left( U(\Delta t;0) U(\Delta t;\tilde \la_1 +\tilde \la_2 ) \right.
\no \no
&&
\Sp \left. U(\Delta t; \tilde \la_1) U(\Delta t;0) \right) D^\dagger(\la_1) \,\, ,
\nonumber
\eq
where $U(\Delta t;\tilde \la) =D(\tilde \la) U(\Delta t) D^\dagger(\tilde \la)$, 
$\tilde \la_i=\la_{i+1}-\la_i$ and $\la_4=\la_1$.
For integer $N$ It is instructive to simulate numerically the above operator studying its dependence on 
the number of discrete steps used to approximate the continuous procedure.
We start with five kicks. The involved parameters 
are taken to be $T=0.1$ and $\hbar X=1$, with the radius of the circle equal to 1. 
The initial point is taken to be the origin of the complex plane rather than $\la_1$, 
or in other words we do not perform the initial and final displacements by 
$D^\dagger(\la_1)$ and $D(\la_1)$.
In the table below are given the percentage deviations of
the absolute values of the (0,0), (0,1), (1,0), (1,1) elements of the evolution 
operator $U(0,T)$ as functions of the number of displacers used to approximate 
the cyclic evolution. 
These are the relevant elements for the evolution of the 
states in the degenerate eigen-space describing a qubit.
Numerical values were obtained 
for 5, 10, 20 and 26 displacers and the percentage is with respect to the value 
obtained with 100 displacers.
\end{multicols}
\widetext
\begin{center}
\begin{tabular}{|c||c|c|c|c|}                   \hline
   &  5     & 10     & 20     & 26     \\       \hline \hline
\Sp 00 \Sp & \Sp 0.2419 \% \Sp & \Sp 0.0595 \% \Sp & \Sp 0.0149 \% \Sp & \Sp 0.0099 \% \Sp \\
\hline
\Sp 01 \Sp & 0.9119 \% & 0.2260 \% & 0.0558 \% & 0.0186 \% \\       \hline
\Sp 10 \Sp & 0.9119 \% & 0.2260 \% & 0.0558 \% & 0.0186 \% \\       \hline
\Sp 11 \Sp & 1.6763 \% & 0.4061 \% & 0.0760 \% & 0.0269 \% \\       \hline
\end{tabular}
\end{center}
\begin{multicols}{2}
\narrowtext
We see that with 26 displacers the error is of the order of 1 in $10^4$ acceptable 
for quantum computation with error correction. This provides an indication for the 
necessary number of devices needed in order to reproduce faithfully the holonomic 
adiabatic loop. 

In the polygon of Fig. \ref{trig} the displacement from point $\la_i$ to $\la_{i+1}$ 
is performed by the operating device with unitary action $D(\la_{i+1}-\la_i)$. 
This device is naturally affected by an error in the position of the point $\la_{i+1}$. 
As an overall effect this will introduce an error in the area of the enclosed surface
bounded by the loop $C$, as well as an error in the ``matching'' of the initial and 
final points $\la_1$ and $\la_{N+1}$. The enclosed area is the parameter of the unitaries 
interpreted as logical gates, while the matching of the points $\la_1$ and $\la_{N+1}$ is 
required in order to apply the adiabatic theorem for the holonomic interpretation of the 
gates. In both cases the error is zero in the first order (i.e. our model is robust) 
with respect to the error introduced by spanning the loop $C$. 

Another appealing characteristic of the holonomic model is the following. Consider 
$\tau_k$ to be the (kicking) time during which each displacer acts, $T$ the overall period of the 
evolution and $\omega$ the characteristic frequency of our system with 
Hamiltonian $H_0$. Then, in order to apply the adiabatic theorem and the kick method 
the following inequality should to hold
\bq
\tau_k \ll \omega^{-1} \ll T
\nonumber
\eq
This condition is robust against small variations of the parameters $\tau_k$ and $T$.

\section{Conclusions}

In this paper we provided a detailed account of the so-called Holonomic approach
to quantum computation (HQC). While in the standard ``dynamical''
picture the information processing 
is performed by a sequence (network) of logic gates
obtained by turning on and off suitable Hamiltonians
in HQC the idea is to exploit the tools of non-abelian gauge theories
to manipulate quantum information.
The quantum codewords are realized by $n$-fold degenerate eigenstates
of an iso-degenerate $d$-parametric family of Hamiltonians 
${\cal F}=\{H(\lambda)\}_\lambda.$
The quantum gates are then given by the holonomies $U(\gamma)\in U(n)$
produced by moving along loops in the manifold $\cal M$ of control parameters $\lambda.$
The holonomies represent the non-commutative generalization of the well-known
Berry geometric phases and their existence is due to the non-trivial geometry
of the set of the computational eigen-spaces thought of as a complex vector 
bundle over the control
manifold ${\cal M}.$

We focused on the case in which these loops are traveled in an adiabatic way.
In this situation one can find explicit expression for the non-abelian ($u(n)$-valued)
connection i.e.,
gauge potential, form
$A$ such that $U(\gamma)= {\bf {P}}\exp \int_\gamma A.$
The computational power of the given connection $A$ is described in terms
of the associated curvature form $F:$
when $d\,(d-1)/2$ components of $F$ span the whole $u(n)$
then one can perform universal QC over the code.
We thoroughly discussed several examples showing explicitly how to design
loops, in the relevant control manifolds, in such a way as to get universal set of gates. 
We argued how, by using suitable Holonomic families $\cal F$ acting on 
multi-partite systems, one can achieve efficient computation
taking advantage of quantum entanglement.
We also discussed a potential experimental
demonstration of HQC in a quantum-optical 
set-up, using artificial Holonomic family $\cal F$ 
generated with an iterated ``kicks'' method.

The HQC approach shows 
how many of the notions and techniques developed in (non-abelian) gauge 
theories can find a natural application and interpretation
in the arena of quantum 
information processing. 
This rather unexpected connection between the tools
used for the description of Nature at its fundamental level
and the ideas of Information/Computation theory is conceptually quite
intriguing. In a slogan HQC suggest that information, besides being ``physical''
it can also be ``geometrical''.
On the more concrete side HQC, similarly to the other geometrical/topological approaches
to QC recently emerged, shows inherent fault-tolerant issues.
This latter fact could make HQC appealing even from the implementation point of view,
in spite of the quite demanding requirements it involves.
Individuation of potential candidates for realization of HQC schemes
in the lab is a major challenge for future research.

\section{Acknowledgment}

%\begin{aknowledgements}
The authors thank M. Rasetti and S. Chountasis for useful discussions and critical reading of the manuscript.
%\end{aknowledgements}

%%%%%%%%%%%%%%%%%%%%%%%%%%%%%%%%%%%%%%%%%%%%%%%%%%%%%%%%%%%%%%%%%%%%%%%%%%%

\appendix
\section{Quantum Holonomies}%%%%%%%%%%%%%%%%%%%%%%%%
In this appendix we shall give a mathematical review of the basic formalism concerning 
quantum geometric phases and their non-abelian generalizations.
Although the material discussed below is by now standard (see for example Ref. \cite{SHWI}
on which our presentation strongly relies)
it has been here introduced for the sake of self-completeness of the paper
and for making it more accessible to readers from the field of Quantum Information.
For the necessary (elementary) background of fiber bundles theory
we refer the reader to the book \cite{NAK}.

\subsection{Abelian phases}
In quantum theory the physical states are represented 
by {\em rays} in a separable Hilbert space $\cal H.$
Mathematically this means that (pure) states are in a one-to-one correspondence
with the elements of the {\em projective} space ${\bf{P}}({\cal H}).$
The latter is defined as the quotient space of $\cal H$ with respect to
the equivalence relation $x\sim y\leftrightarrow \exists \lambda \in\CC-\{0\}\,/\,
y=\lambda\,x,\,\forall x,y\in{\cal H}.$
Alternatively one can consider the unit sphere in ${\cal H},$ 
$S^\infty :=\{|\psi\rangle\,/\,\||\psi\rangle\|=1\}$,
factored by the $U(1)$ action $(|\psi\rangle,\,e^{i\theta})\mapsto e^{i\theta}\,|\psi\rangle.$
In this case one writes ${\bf{P}}({\cal H})= S^\infty/U(1).$

The projection 
\begin{equation}
\pi\colon S^\infty\rightarrow {\bf{P}}({\cal H})
\label{princ-bundle}
\end{equation}
defines a $U(1)$ principal bundle over the base space ${\bf{P}}({\cal H})$ 
with total space $S^\infty.$ This fiber bundle is the natural setting in which abelian holonomies
i.e., Berry phases, appear.
The fiber over the point $|\psi\rangle\langle\psi|\in {\bf{P}}({\cal H}) $
is given by $F_{|\psi\rangle}:=\{ e^{i\,\theta}\,|\psi\rangle\,/\,\theta\in [0,\,2\,\pi)\}.$

Any other principal $U(1)$ bundle can be written in terms of (\ref{princ-bundle}) as follows.
Suppose $\Phi$ to be a (smooth) map from the parameter 
manifold ${\cal M}$ in ${\bf{P}}({\cal H}),$
then one can construct the {\em pull-back} bundle $\Phi^* S^\infty,$ with total
space $\bigcup_{\lambda\in{\cal M}} F_{\Phi(\lambda)}$ and projection 
$\pi^\Phi\colon F_{\Phi(\lambda)} \rightarrow \lambda.$
A map $\lambda\mapsto |\psi_\lambda\rangle \in F_{\Phi(\lambda)}$ is a {\em
section } of the bundle $\Phi^* S^\infty.$

The Schr\"odinger equation $i\,\partial/\partial t \,|\psi\rangle= H\,|\psi\rangle$
determines a temporal evolution $t\mapsto |\psi(t)\rangle$ in the total space $S^\infty$
and via $\pi$ in ${\bf{P}}({\cal H}).$
Moving the other way round i.e., from an evolution in ${\bf{P}}({\cal H})$
and obtaining an evolution in $S^\infty,$
requires the introduction of a new key ingredient: a {\em connection}.

The $U(1)$ action over $S^\infty$ defines the {\em vertical} direction
at each point along the fiber. The connection $u(1)$-valued $1$-form $A$ allows
to define the {\em horizontal} direction as well.
Once this field of directions is given one can realize the horizontal {\em lift}
of any curve in the base. 

To build the connection let us start by observing
that $S^\infty$ inherits from ${\cal H}$ a natural hermitian structure.
Such structure provides a natural mean for defining the horizontal directions
at each point: 
the ones orthogonal to the fiber i.e., to $|\psi\rangle.$ 
Given the curve $t\mapsto |\psi(t)\rangle\in S^\infty$ we decompose
the tangent vector $|\dot\psi\rangle:=d|\psi(t)\rangle/dt$ 
as follows $|\dot\psi\rangle= \langle\psi|\dot\psi\rangle\,|\psi\rangle + |h_\psi\rangle.$ 
Where the horizontal component $|h_\psi\rangle$ satisfies the relation 
$\langle\psi|h_\psi\rangle=0.$

The connection can be evaluated explicitly by splitting the operators $d/dt$
according its vertical and horizontal components: 
${d}/{dt} =\alpha\,{ \partial}/{\partial\theta} + \sum_\mu B^\mu\, D_\mu$,
where $\mu$ label the local chart coordinates $(\lambda_\mu)$ of the base manifold $\cal M.$
The horizontal operators $D_\mu$ are referred to as the {\em covariant} derivatives,
they are given by
$D_\mu:= {\partial}/{\partial\lambda_\mu}+ A_\mu \,{\partial}/
{\partial\theta}.$
The $u(1)$-valued $1$-form $A:=\sum_\mu A_\mu \,d\lambda_\mu$ is the connection form.
Applying $d/dt$ to $|\psi(t)\rangle$ from the horizontal part one gets
$\langle\psi|\,D_\mu\,|\psi\rangle=
\langle\psi|\,\partial/\partial\lambda_\mu +\,A_\mu\,\partial/\partial\theta\,|\psi\rangle=0,$
from which $A_\mu \,\langle\psi|\partial/\partial\theta\,|\psi\rangle= -
\langle\psi|\,\partial/\partial\lambda_\mu\,|\psi\rangle$.
Since $\partial/\partial\theta\,|\psi\rangle=i\,|\psi\rangle$ one obtains
\begin{equation}
A_\mu= i\,\langle\psi|\frac{\partial}{\partial\lambda_\mu}\,|\psi\rangle \,\, .
\label{berry-conn}
\end{equation}
A section $s\colon {\cal M}\mapsto S^\infty\,/\,
\lambda\mapsto|\psi\rangle_\lambda$ of (\ref{princ-bundle}),
under a local coordinate change, transforms as $|\psi\rangle_\lambda\mapsto \exp(i\,\theta(\lambda))
\,|\psi\rangle_\lambda$ (gauge transformation).
From this relation and eq. (\ref{berry-conn}) one can check that the connection form
changes according to $A_\mu\mapsto A_\mu -\partial\theta/\partial\lambda_\mu.$

Let us consider now a loop 
$\gamma\colon[0,\,T]\mapsto {\bf{P}}({\cal H})$
in the base space.
Using the connection form (\ref{berry-conn}) we can lift it to the total space
$S^\infty,$ let $|\tilde\psi(t)\rangle$ be such an horizontal lift.
One can write $|\tilde\psi(t)\rangle= e^{i\,f(t)}\,|\psi(t)\rangle,$
where $|\psi(t)\rangle= (s\circ\gamma)(t),$ is a {\em closed} in $S^\infty$
obtained composing $\gamma$ with a section.
One has $|\tilde\psi(T)\rangle= e^{i\,\beta}\,|\tilde\psi(0)\rangle,$
where $\beta:= f(T)-f(0).$
From the above relations and the horizontality condition 
$\langle \tilde\psi(t)|\dot{\tilde\psi(t)}\rangle=0,$ one gets
$\beta= \int_0^T dt \langle\psi(t)|\dot\psi(t)\rangle.$
Since $|\dot\psi\rangle=(\dot\theta\partial/\partial\theta 
+\sum_\mu \dot\lambda_\mu \partial/\partial\lambda_\mu)\,|\psi\rangle,$
and $\theta(T)=\theta(0)$ mod $2\,\pi,$ by using Eq. (\ref{berry-conn})
one finds 
\begin{equation}
\beta= \sum_\mu \int_\gamma d\lambda_\mu A_\mu =\int_\gamma A.
\label{hol}
\end{equation}
This expression of the $U(1)$-holonomy
is manifestly gauge invariant and it does not depend on the Hamiltonian:
it is of a pure {\em geometrical} origin.

In general the state vector is {\em not} horizontal 
$\langle\psi|\dot\psi\rangle=-i\,\langle\psi|H(t)|\psi\rangle\neq 0,$
in this case besides the geometrical term (\ref{hol})
one has a {\em dynamical} contribution to the phase
given by $\alpha:=-\int_0^T \langle\psi(t)|H(t)|\psi(t)\rangle.$
At variance with the former this latter term depends
on the Hamiltonian, moreover one can get rid of it
by a gauge transformation $|\psi(t)\rangle\mapsto U(t)\,|\psi(t)\rangle,$ 
where
\bq
U(t):=\exp [i\int_0^t d\tau\,\langle\psi(\tau)|H(\tau)|\psi(\tau)\rangle].
\nonumber
\eq
By introducing the {\em curvature} $2$-form $F:=dA=\sum_{\mu\nu} F_{\mu\nu} d\lambda_\mu\wedge d\lambda_\nu,\,
F_{\mu\nu}=\partial A_\mu/\partial\lambda_\nu-\partial A_\nu/\partial\lambda_\mu,$
and using Stokes theorem, eq. (\ref{hol}) can be rewritten as
$\beta:= \int_\Sigma F$ where $\Sigma$ is any surface having $\gamma([0,\,T])$ as a boundary.
A non-vanishing $F$ describes a non-trivial geometry of the bundle (\ref{princ-bundle}).
One can generalize this construction to the more complex case in which the fiber
is a $n$-dimensional complex space over which acts the group $U(n).$
This {\em non-Abelian} situation will be now reviewed.

\subsection{Non Abelian generalization}%%%%%%%%%%%%%%%%%%

Let us consider a family of Hamiltonians $\{ H(\lambda) \}_{\lambda\in {\cal M}}$
where ${\cal M}$ is the control manifold. We shall consider loops $\gamma$ in ${\cal M},$ 
and define $H(t):=H_{\gamma(t)},\,t\in[0,\,T].$ 
In general $H(\lambda)=H_{\gamma(t)}$ has $R$ different eigenvalues $\{ \varepsilon_l\}_{l=
1}^R$ with degeneracies $\{n_l(\lambda)\}.$
We shall assume that {\em no} level crossing occur i.e., $n_l(\lambda)=n_l.$
Let $\Pi_l(\lambda)$ denote the projector over the eigen-space
${\cal H}_i(\lambda) :=\mbox{span}\,\{|\psi^\alpha_{l}(\lambda)\rangle\}_{\alpha=1}^{n_l},$
of $H(\lambda)$. The spectral $\lambda$-dependent resolution of the Hamiltonians is then
$ H(\lambda)=\sum_{l=1}^R \varepsilon_l(\lambda) \, \Pi_l(\lambda)$.
The mapping $\lambda\mapsto |\psi^\alpha_{l}(\lambda)\rangle$ 
defines a section of the bundle 
\begin{equation}
\frac{U(N)}{U(N-n_i)}\rightarrow \frac{U(N)}{U(n_i)\times U(N-n_i)}
\label{stief-grass}
\end{equation}
where $N:=\mbox{dim}\,{\cal H}.$
The total (base) space of the $U(n_l)$-principal bundle (\ref{stief-grass}) is known as
the Stiefel (Grassmann) manifold and it is denoted by $V_{N, n_l}$ ($G_{N, n_l}$).
In the very same way as for the abelian case discussed above the hermitian structure
over ${\cal H}$ provides a natural notion of horizontality:
tangent vectors are horizontal if they are orthogonal to the fiber.
Notice that one recovers the abelian (${\cal H}\cong \CC^N$) case by setting $n_l=1,$
indeed $G_{N,1}={\bf{CP}}^{N-1}={\bf{P}}({\cal H}).$

The state vector evolves according the time-dependent Schr\"odinger equation
$i\,\partial_t|\psi(t)\rangle= H_{\gamma(t)}\,|\psi(t)\rangle.$
In the case in which the loop $\gamma$'s are traveled sufficiently slow
one avoids transitions among different energy levels. In this {\em adiabatic} limit
any initial preparation $|\psi_0\rangle\in{\cal H}$ belonging to 
some energy eigen-space ${\cal H}_l$ will be mapped, after the period $T,$ onto:
$|\psi(T)\rangle=U_\gamma\,|\psi_0\rangle\in{\cal H}_l.$
For the sake of concreteness let us focus on the eigen-space associated with the 
eigenvalue $\varepsilon_l=0$. Also let $\{|\psi_\alpha(t)\rangle\}$ be the 
corresponding orthonormal basis at the instant $t.$

Let $\eta_\alpha(t)=\sum_\beta U_{\alpha\beta}(t)\, \psi_\beta(t)$ be
the solution of the Schr\"odinger equation with initial condition 
$\eta_\alpha(0)=\psi_\alpha.$
By imposing, at each instant,
the orthogonality i.e., horizontality, 
conditions $\langle\eta_\beta|\eta_\alpha\rangle=\delta_{\alpha\beta}$
one gets by differentiation
\begin{eqnarray}
0&=&\langle\eta_\beta|\dot\eta_\alpha\rangle=\sum_\delta (
\dot U_{\alpha\delta}\,\langle\eta_\beta|\psi_\delta\rangle
+ U_{\alpha\delta}\, \langle\eta_\beta|\dot\psi_\delta\rangle)
\nonumber \\
&=&\dot U_{\alpha\delta}\, U_{\beta\delta}^*+\sum_{\delta\tau} U_{\alpha\delta}\, U^*_{\beta\tau}\,
\langle\psi_\tau|\dot\psi_\delta\rangle,
\end{eqnarray}
from which follows $(U^{-1}\,\dot U)_{\beta\alpha}= A_{\alpha\beta}$
where we have defined $A_{\alpha\beta}:=\langle\psi_\beta|\dot\psi_\alpha\rangle.$
From the above relations it follows that
\begin{equation}
U(t)={\bf{T}}\exp\int_0^T d\tau A(\tau)={\bf{P}}\exp\int_\gamma A,
\end{equation}
where the $u(n)$-valued $1$-form $A=\sum_\mu A_\mu d\lambda_\mu $
is given by $(A_\mu)_{\alpha\beta}= \langle\psi_\alpha|\partial/\partial\lambda_\mu|\psi_\beta\rangle.$
Note that the matrix character of the connection $A$ demands for
the path ordering ${\bf P}$ to take place. For a reparametrization of the
loop $\gamma$ in terms of a variable $x \in [a,b]$ and for $\A(x)=A_\mu {d \la_\mu \over dx}$ 
it is formally defined by
\bq
\Gamma_A(\gamma) \equiv {\bf P} e^{\int _a^b \A(x) dx } =\sum _{n=0} ^{\infty} {1 \over n! } 
{\bf P} \left( \int _a^b \A(x) dx \right)^n
\nonumber
\eq
with
\bq
&& \Sp
{1 \over n!} {\bf P} \left( \int _a^b \A(x) dx \right)^n :=
\no \no
&&
\int _a^b dx_1 \int _a ^{x_1} dx_2 ... \int _a ^{x_{n-1}} dx_n \A(x_1) \A(x_2)... \A(x_n)
\label{order}
\eq
%\be
%{1 \over n!} {\bf P} \left( \int _a^b \A(x) dx \right)^n :=
%\int _a^b dx_1 \int _a ^{x_1} dx_2 ... \int _a ^{x_{n-1}} dx_n \A(x_1) \A(x_2)... \A(x_n)
%\label{order}
%\ee
%
%\begin{multicols}{2}

where $x_k=a+{k \over n}(b-a)$. 

%%%%%%%%%%%%%%%%%%%%%%%%%%
%%%%%%%%%%%%%%%%%%%%%%%%%%%%%%%
\end{multicols}
\end{document}